%% file: main.tex
\documentclass[acmsmall,nonacm]{acmart}

\usepackage{multirow}
\usepackage{dirtytalk}
\usepackage{amsmath}

\usepackage{amssymb}
\usepackage{amsthm}
\usepackage{mathtools}

\usepackage{hyperref}
\usepackage[capitalize,noabbrev]{cleveref} %
\usepackage{url}
\usepackage{microtype}
\usepackage{relsize}
\usepackage{natbib}

\usepackage{algorithm}
\usepackage{algorithmicx}
\usepackage{algpseudocode}
\usepackage{arydshln}

\usepackage{listings}
\usepackage{xspace}
\usepackage{multirow}
\usepackage{thmtools}
\usepackage{thm-restate}
\usepackage{breqn}
\usepackage{subfig}

\usepackage{booktabs}       %
\usepackage{nicefrac}       %
\usepackage{microtype}      %
\usepackage{colortbl}       %
\usepackage{stackengine}
\usepackage{lipsum}
\usepackage{dsfont}
\usepackage{graphicx}
\usepackage{tabularx}
\usepackage{float}
\usepackage{mathtools}
\usepackage{mathrsfs}
\usepackage{enumitem}
\usepackage{xspace}
\usepackage{xparse}
\usepackage[export]{adjustbox}

\usepackage{conf/tikzit}
\input{conf/global.tikzstyles}

\usepackage[framemethod=TikZ]{mdframed}
\usepackage{footnote}
\usepackage{booktabs}
\usepackage{soul}
\AtBeginDocument{%
  }

\setcopyright{acmlicensed}
\copyrightyear{2024}
\acmYear{2024}
\acmDOI{XXXXXXX.XXXXXXX}

\acmConference[Conference acronym 'XX]{Make sure to enter the correct
  conference title from your rights confirmation emai}{June 03--05,
  2018}{Woodstock, NY}
\acmISBN{978-1-4503-XXXX-X/18/06}

\input{conf/defs}

\input{conf/confs}

\input{conf/macros-l}

\begin{document}

\title{The Correlated Gaussian Sparse Histogram Mechanism}

\author{Christian Janos Lebeda}
\authornote{All authors contributed equally to this research.}
\orcid{xxx}
\affiliation{%
  \institution{Inria, University of Montpellier}
  \country{France}
}
\email{christian.j.lebeda@gmail.com}

\author{Lukas Retschmeier}
\authornotemark[1]
\affiliation{%
  \institution{BARC, University of Copenhagen}
  \country{Denmark}
}
\email{lure@di.ku.dk}

\renewcommand{\shortauthors}{Lebeda, Retschmeier}

\include{content/abstract}

\received{Dec 2025}

\maketitle

\input{content/introduction}

\input{content/preliminaries}
\input{content/gshm}

\input{content/results-intro}

\input{content/top-k}

\input{content/empirical-eval}
\input{content/aggregators}

\input{content/discrete-gaussian}

\input{content/related}
\input{content/open}

\input{content/ack}
\bibliographystyle{ACM-Reference-Format}
\bibliography{references.bib}

\appendix
\newpage
\input{content/appendix}

\end{document}

%% file: conf/global.tikzstyles
\tikzstyle{BLACK}=[draw=black, shape=circle, fill=black, inner sep=3pt]
\tikzstyle{DOM}=[fill={rgb,255: red,122; green,0; blue,42}, draw=black, shape=circle, inner sep=3pt]
\tikzstyle{NONE}=[fill={rgb,255: red,247; green,33; blue,212}, draw=black, shape=circle, inner sep=3pt]
\tikzstyle{NTWO}=[fill={rgb,255: red,0; green,177; blue,219}, draw=black, shape=circle, inner sep=3pt]
\tikzstyle{NTHR}=[fill={rgb,255: red,140; green,143; blue,14}, draw=black, shape=circle, inner sep=3pt]
\tikzstyle{GRAYN}=[fill={rgb,255: red,128; green,128; blue,128}, draw=black, shape=circle, inner sep=3pt]
\tikzstyle{BRGRAY}=[fill={rgb,255: red,236; green,236; blue,236}, draw=none, shape=circle, inner sep=3pt]
\tikzstyle{TEXTSTD}=[fill=white, draw=black, shape=rectangle, tikzit shape=rectangle, text width=3cm, rounded corners]
\tikzstyle{TEXTHIGH}=[fill={rgb,255: red,231; green,245; blue,255}, draw=black, shape=rectangle, tikzit shape=rectangle, text width=3cm, rounded corners]
\tikzstyle{TEXTHIGHGRAY}=[fill={rgb,255: red,220; green,220; blue,220}, draw=black, shape=rectangle, tikzit shape=rectangle, text width=3cm, rounded corners]
\tikzstyle{TEXTHIGHGREEN}=[fill={rgb,255: red,149; green,242; blue,145}, draw=black, shape=rectangle, tikzit shape=rectangle, text width=3.4cm, rounded corners]
\tikzstyle{TEXTHIGHYELLOW}=[fill={rgb,255: red,250; green,250; blue,160}, draw=black, shape=rectangle, tikzit shape=rectangle, text width=3cm, rounded corners]

\tikzstyle{EDGE}=[-, fill=none, line width=2pt]
\tikzstyle{BLUE}=[-, draw={rgb,255: red,0; green,101; blue,189}]
\tikzstyle{GRAY}=[-, fill={rgb,255: red,128; green,128; blue,128}]
\tikzstyle{DARKGREEN}=[-, fill=none, draw={rgb,255: red,19; green,100; blue,13}, line width=1.25pt]
\tikzstyle{big dash}=[-, dashed, dash pattern=on 4mm off 2mm, fill={rgb,255: red,178; green,255; blue,253}]
\tikzstyle{big dash thick}=[-, thick, dashed, dash pattern=on 4mm off 2mm, fill={rgb,255: red,178; green,255; blue,253}]
\tikzstyle{new edge style 0}=[-, fill={rgb,255: red,70; green,255; blue,243}, line width=1.5pt]
\tikzstyle{FILLGREEN}=[-, fill={rgb,255: red,144; green,237; blue,134}, line width=1pt]
\tikzstyle{FILLRED}=[-, fill={rgb,255: red,250; green,128; blue,114}, line width=1pt]
\tikzstyle{FILLBLUE}=[-, fill={rgb,255: red,162; green,232; blue,230}, line width=1pt]
\tikzstyle{FILLPURPLE}=[-, fill={rgb,255: red,255; green,233; blue,239}]
\tikzstyle{FILLDARKBLUE}=[-, fill={rgb,255: red,238; green,230; blue,255}]
\tikzstyle{FILLDARKBLUEINVISIBLE}=[-, draw=none, fill={rgb,255: red,238; green,230; blue,255}]
\tikzstyle{FILLGREENINVISBLE}=[-, draw=none, fill={rgb,255: red,232; green,255; blue,207}]
\tikzstyle{ULTRAGRAY}=[-, draw={rgb,255: red,150; green,150; blue,150}]
\tikzstyle{EDGEDASHED}=[-, dashed, fill=none, line width=1pt]
\tikzstyle{simple directed edge}=[->, draw=black, thick, line width=0.75mm]
\tikzstyle{simple dashed directed edge}=[->, draw=black, dashed]
\tikzstyle{SLIGHTLYDASHED}=[-, dotted, fill=none, draw={rgb,255: red,128; green,128; blue,128}]
\tikzstyle{simple dashed directed edge}=[->, draw=black, line width=0.5mm, dashed]
\tikzstyle{simple dashed directed edge not thick}=[->, draw=black, line width=0.45mm, dashed]
\tikzstyle{simple}=[->, draw=black, thick]
\tikzstyle{noise arrow}=[draw={rgb,255: red,128; green,44; blue,127}, fill=none, ->, line width=0.4mm]
\tikzstyle{invis}=[-, draw=none]
\tikzstyle{red-edge}=[-, draw={rgb,255: red,191; green,0; blue,64}, fill=none, line width=1.5pt]
\tikzstyle{thick}=[-, fill=none, line width=1.25pt]
\tikzstyle{noise arrow stopper}=[-, line width=0.35mm, draw={rgb,255: red,116; green,66; blue,128}]
\tikzstyle{black dotted}=[-, dotted, line width=1pt]
\tikzstyle{new edge style 1}=[line width=0.5mm, draw=black, ->, fill=none]

%% file: conf/defs.tex
\renewcommand{\epsilon}{\varepsilon}

\newcommand{\deltaone}{\ensuremath{\delta_{\text{\footnotesize{gauss}}}}}
\newcommand{\deltatwo}{\ensuremath{\delta_{\text{\footnotesize{inf}}}}}

\newcommand{\X}{\ensuremath{\vec{X}}}
\newcommand{\Nd}{\ensuremath{\mathcal{N}}}
\newcommand{\discGauss}{\ensuremath{\mathcal{N}_{\mathbb{Z}}}}
\newcommand{\algoname}{CSH}
\newcommand{\atd}{\emph{add-the-deltas}}

\setul{}{1.5pt} %
\definecolor{sbOrange}{HTML}{D78355}
\definecolor{sbGreen}{HTML}{60AD72}
\definecolor{sbRed}{HTML}{C7595C}
\definecolor{sbBlue}{HTML}{4C72B0}

    \definecolor{nicepurple}{HTML}{802c7f}
    \definecolor{orangeish}{HTML}{f1a340} %
    \definecolor{grayish}{HTML}{f7f7f7} %
    \definecolor{purpleish}{HTML}{998ec3} %
    \definecolor{blueish}{HTML}{004488}
    \definecolor{darkgreen}{RGB}{2,100,64} %
    \definecolor{lightgreen}{HTML}{b2f2bb}
    \definecolor{lightblueish}{RGB}{230,244,255}
\PassOptionsToPackage{table,xcdraw,svgnames,dvipsnames}{xcolor}

\newtheorem{theorem}{Theorem}[section]
\newtheorem{lemma}[theorem]{Lemma}

\newtheorem{definition}[theorem]{Definition}

\newtheorem{corollary}[theorem]{Corollary}

\renewenvironment{quote}
  {\list{}{\rightmargin=0.3cm \leftmargin=0.3cm}%
   \item\relax}
  {\endlist}

%% file: conf/confs.tex
\DeclareCaptionFormat{custom}
{%
    \textbf{#1#2}\textit{\small #3}
}

%% file: conf/macros-l.tex
\newcommand{\R}{\ensuremath{\mathbb{R}}}
\newcommand{\N}{\ensuremath{\mathbb{N}}}

\crefname{lstlisting}{Listing}{Listings}
\newcommand{\basicCodeStyle}{\ttfamily\small}
\newcommand{\keywordCodeStyle}{\bfseries}
\newcommand{\builtInCodeStyle}{\itshape\color{orangeish}}
\newcommand{\typesCodeStyle}{\ttfamily\small\color{blueish}}

\lstdefinelanguage{plan}{
	classoffset=0,
	morekeywords={True, False, return, null, if, in, while, do, else, case, break, continue, for, def, log, sqrt, norm},
		keywordstyle=\keywordCodeStyle,
	classoffset=1,
	morekeywords={privQuant, recenter,  scale, clip, noise},
	keywordstyle=\typesCodeStyle,
	classoffset=2,
	morekeywords={estimateStd, rankError, divideBudget, size, scope, center},
	keywordstyle=\builtInCodeStyle,
	classoffset=0,
	morecomment=[s]{/*}{*/},
	morecomment=[l]//,
	morestring=[b]",
	morestring=[b]'
}

\renewcommand\vec{\mathbf}

\newcommand{\lukas}[1]{\textcolor{darkgreen}{[Lukas: #1]}}
\newcommand{\christian}[1]{\textcolor{orange}{[Christian: #1]}}

\newcommand{\cA}{\mathcal{A}}

%% file: content/abstract.tex
\begin{abstract}
We consider the problem of releasing a sparse histogram under \((\varepsilon, \delta)\)-differential privacy.
The stability histogram independently adds noise from a Laplace or Gaussian distribution to the non-zero entries and removes those noisy counts below a threshold. %
Thereby, the introduction of new non-zero values between neighboring histograms is only revealed with probability at most \(\delta\), and typically, the value of the threshold dominates the error of the mechanism.
We consider the variant of the stability histogram with Gaussian noise.

\noindent Recent works ([Joseph~and~Yu,~COLT~'24] and [Lebeda,~SOSA~'25]) reduced the error for private histograms 
using correlated Gaussian noise.
However, these techniques can not be directly applied in the very sparse setting.
Instead, we adopt Lebeda's technique and show that adding correlated noise to the non-zero counts only allows us to reduce the magnitude of noise when we have a sparsity bound.
This, in turn, allows us to use a lower threshold by up to a factor of \(1/2\) compared to the non-correlated noise mechanism.
We then extend our mechanism to a setting without a known bound on sparsity. 
Additionally, we show that correlated noise can give a similar improvement for the more practical discrete Gaussian mechanism.
\end{abstract}

%% file: content/introduction.tex
\section{Introduction}
Releasing approximate counts is a common task in differential private data analysis.
For example, the task of releasing a search engine log, where one wants to differentially private release the number of users who have searched each possible search term.

The full domain of search terms is too large to work with directly, but the vector of all search counts is extremely sparse as most phrases are never searched.
The standard approach to exploit sparsity while ensuring differential privacy is the \emph{stability histogram} %
\cite{BalcerVadhan-DP-finite-computers, Korolova09-DP-approx-sparse-hist, BunNS16-simultaneous-learning-multiple-concepts,ALP,googlelibthreshold,Gotz2012,WilsonZLDSG20-SQL-bounded-contribution}.  
The idea is to preserve sparsity by filtering out zero counts and then adding noise from a suitable probability distribution to the remaining ones.
Unfortunately, some filtered-out counts could be non-zero in a neighboring dataset, and thus, revealing any of them would violate privacy.
To limit this privacy violation, the stability histogram introduces a threshold $\tau$ and releases only those noisy counts that exceed $1 + \tau$.
This way, we might still reveal the true dataset among a pair of neighboring datasets if one of these counters exceeds the threshold, but with an appropriate choice of $\tau$, this is extremely unlikely to happen.

In our setting, one single user may contribute to many counts; hence, using the \emph{Gaussian Sparse Histogram Mechanism} (GSHM) \cite{googlelibthreshold,Wilkins24-GaussianSparseHistogramMechanism} might be preferable. 
The GSHM is similar to the stability histogram, but it replaces noise from the Laplace distribution with Gaussian noise.  
To analyze the $(\epsilon, \delta)$ guarantees of the GSHM, one has to find an upper bound for $\delta$, which is influenced by the standard Gaussian mechanism and the small probability of infinite privacy loss that can happen when the noise exceeds $\tau$.
We denote these two quantities as $\deltaone$ and $\deltatwo$.
Recently, \citet{Wilkins24-GaussianSparseHistogramMechanism} gave exact privacy guarantees of the GSHM, which improves over the analysis by \citet{googlelibthreshold} where $\deltaone$ and $\deltatwo$ were simply added together.
Their main contribution is a more intricate case distinction of a single user's impact on the values of $\deltaone$ and $\deltatwo$.
This allows them to use a lower threshold with the same privacy parameters.
Their improvement lies in the constants, but it can be significant in practical applications because the tighter analysis essentially gives better utility at no privacy cost.

Our goal is to reduce the error further.
Because the previous analysis is exact, we must exploit some additional structure of the problem.
Consider a $d$-dimensional histogram $H(\X) = \sum_{i = 1}^nX_i$ for $\X = (X_1, ... X_n)$ for users with data $X_i \in \{0,1\}^d$.
Notice that \emph{adding} a single user to $\X$ can only increase counts in $H(\X)$, whereas \emph{removing} one can only decrease counts.
\citet{lebeda2024} recently showed how to exploit this monotonicity by adding a small amount of correlated noise.
This reduces the total magnitude of noise almost by a factor of $2$ compared to the standard Gaussian Mechanism.
So, it is natural to ask if we can adapt this technique to the setting when the histogram is sparse.
This motivates our main research question: %

\begin{quote}
{\bf Question: } Can we take advantage of monotonicity and use correlated noise to improve the \emph{Gaussian Sparse Histogram Mechanism}?
\end{quote}

\subsection{Our Contribution}

We answer this question in the affirmative and introduce the \emph{Correlated Stability Histogram} (\algoname).
Building on the work by \citet{lebeda2024}, we extend their framework to the setting of releasing a histogram under a sparsity constraint, that is $\|H(\X)\|_0 \leq k$ for some known $k$. 
This is a natural setting that occurs e.g. for Misra-Gries sketches, where $k$ is the size of the sketch \cite{MisraGries82,LebedaTetek23-DPMG}. 

\vspace{1.5mm}
\noindent \textbf{Correlated GSHM. \,} 
We introduce the \emph{Correlated Stability Histogram} (CSH), a variant of the GSHM using correlated noise. 
Our algorithm achieves a better utility-privacy trade-off than GSHM for $k$-sparse histograms.
Similar to the result of \citet{lebeda2024}, we show that correlated noise can reduce the error by almost a factor of $2$ at no additional privacy cost.
In particular, our main result lies in the following (informally stated) theorem:

\begin{theorem}[The Correlated Stability Histogram (Informal)]
Let $H(\X) = \sum_i^n X_i$ denote a histogram with bounded sparsity, where $\X = (X_1, ... X_n)$ and $X_i \in \{0, 1\}^d$. %
If the GSHM privately releases $H(\X)$ under $(\epsilon, \delta)$-DP with noise magnitude $\sigma$ and removes noisy counters below a threshold $\tau$,
then the \emph{Correlated Stability Histogram} (\algoname) releases $H(\X)$ also under $(\epsilon, \delta)$-DP with noise of magnitude ${\sigma}/{2} + o(1)$ and removes noisy counters below a threshold ${\tau}/{2} + o(1)$.
\end{theorem}

As a baseline, we first analyze our mechanism using the \atd~ \cite{googlelibthreshold} approach and show that even this approach outperforms the exact analysis in \cite{WilsonZLDSG20-SQL-bounded-contribution} in many settings.
We then turn our attention to a tighter analysis, which uses an intricate case distinction to upper bound the parameter $\delta$.
The proof by \citet{Wilkins24-GaussianSparseHistogramMechanism} does not carry directly over because of the increased complexity due to the correlated noise sample.
Furthermore, we complement our approach with the following results:

\vspace{1.5mm}
\noindent \textbf{Generalization \& Extensions. \,}
We extend our mechanism to multiple other settings, including extensions considered in \citet{Wilkins24-GaussianSparseHistogramMechanism}.
First, we generalize our approach to allow an additional threshold 
that filters out infrequent data in a pre-processing step.
Second, we discuss how other aggregate database queries can be included in our mechanism.
Last, we generalize to top-$k$ counting queries when we have no bound for $\|H(\X)\|_0$.

\vspace{1.5mm}
\noindent \textbf{Discrete Gaussian Noise. \,}
Our mechanism achieves the same improvement over GSHM when noise is sampled from the discrete Gaussian rather than the continuous distribution.
We present a simple modification to make our mechanism compatible with discrete noise. 
This has practical relevance even for the dense setting considered by \citet{lebeda2024}.

\vspace{1.75mm}
\noindent Additionally, \emph{Experimental} evaluations support our claim that our approach can improve utility by up to a factor of $2$ compared with the standard GSHM. %

\vspace{0.65em} %
\noindent{\bf\bf  Organization.}
The rest of the paper is organized as follows.
\cref{sec:background} introduces the problem formally and reviews the required background.
\cref{sec:algorithm} introduces the \emph{Correlated Stability Histogram}, our main algorithmic contribution for the sparse case.
Furthermore, in \cref{sec:top-k} we generalize our approach to the non-sparse setting.
The experiments in \cref{sec:empirical-eval} confirm our claim that we improve the error over previous techniques.
The extensions of our technique to match the setting of \citet{Wilkins24-GaussianSparseHistogramMechanism} are discussed in \cref{sec:aggregators}.
In \Cref{sec:discrete}, we show how to adapt our mechanism for discrete Gaussian noise.
Finally, we conclude the paper in \Cref{sec:conclusion} and discuss open problems for future work.

%% file: content/preliminaries.tex
\section{Preliminaries and Background}\label{sec:background}

Given a dataset $\X \in \mathcal{U}^{\mathbb{N}}$, we want to perform an aggregate query $\cA$ under differential privacy.
We focus on settings where each user %
has a set of elements, and we want to estimate the count of all elements over the dataset.
Therefore, we consider a dataset $\X=(X_1,\dots,X_n)$ of $n$ data points where each $X_i \in \{0, 1\}^d$.
Our goal is to output a private estimate of the histogram $H(\X) \in \mathbb{N}^d$, %
where $H(\X) = \sum_{i = 1}^n X_i$.
For some vector $\vec{H}\in \R^d$, we define the \emph{support} as $U(\vec{H}) = \{i \in [d]: H_i \neq 0 \}$
and denote the $\ell_0$ norm as $\| \vec{H} \|_0 \coloneqq \vert U(\vec{H}) \vert$, being the number of non-zeroes in $\vec{H}$.
In this work, we focus on settings where the dimension $d$ is very large or infinite.

Proposed by~\citet{dwork06calibrating}, differential privacy (DP) is a property of a randomized mechanism.
The intuition behind differential privacy is that privacy is preserved by ensuring that the output distribution does not depend much on any individual's data.
In this paper, we consider $(\varepsilon, \delta)$-differential privacy (sometimes referred to as approximate differential privacy) together with the add-remove variant of neighboring datasets as defined below.
Note that by this definition $|\X| = n$ and $|\X'| = n \pm 1$:

\begin{definition}[Neighboring datasets]
    \label{def:neighboring-datasets}
    A pair of datasets are neighboring, denoted ${\X \sim \X'}$, if there exists an $i$ such that either ${\X = (X'_1, \dots, X'_{i - 1}, X'_{i + 1}, \dots, X'_{n + 1})}$ or $\X' = (X_1, \dots, X_{i - 1}, X_{i + 1}, \dots, X_{n})$ holds.
\end{definition}

\begin{definition}[{\cite{dworkRothBook}} $(\varepsilon, \delta)$-differential privacy]
    \label{def:differential-privacy}
Given $\epsilon$ and $\delta$, a randomized mechanism $\mathcal{M}: \mathcal{U}^{\mathbb{N}} \rightarrow \mathcal{Y}$ satisfies $(\epsilon, \delta)$-DP, if for every pair of neighboring datasets $\X \sim \X'$ and every measurable set of outputs $Y \in \mathcal{Y}$ it holds that
\[
    \Pr[\mathcal{M}(\X) \in Y] \leq e^\epsilon \Pr[\mathcal{M}(\X') \in Y] + \delta \,.
\]
Differential privacy is immune to post-processing:
Let $\mathcal{M}:\mathcal{U}^{\mathbb{N}} \rightarrow R$ be a randomized algorithm that is $(\epsilon, \delta)$-DP. Let $f: R \rightarrow R'$ be an arbitrary randomized mapping. 
Then $f\circ \mathcal{M}:\mathcal{U}^{\N} \rightarrow R'$ is $(\epsilon, \delta)$-DP.
\end{definition}

\noindent An important concept in differential privacy is the sensitivity of a query, which restricts the difference between the output for any pair of neighboring datasets.
We consider both the $\ell_2$ sensitivity and, more generally, the sensitivity space of the queries in this paper.

\begin{definition}[Sensitivity space and $\ell_2$ sensitivity]
    \label{def:sensitivity}
    The sensitivity space of a deterministic function $f : \mathcal{U}^{\mathbb{N}} \rightarrow \mathbb{R}^d$ is the set $\Delta f = \{f(\X) - f(\X') \in \mathbb{R}^d \mid \X \sim \X' \}$.
    The $\ell_2$ sensitivity of $f$ is defined as
    \[
        \Delta_2 f = \max_{\X \sim \X'} \|f(\X) - f(\X')\|_2 = \max_{x \in \Delta f} \| x \|_2 \,,
    \]    
    where $\| \vec{x} \|_2 = \sqrt{\sum_{i = 1}^d x_i^2}$ denotes the $\ell_2$ norm of any $\vec{x} \in \mathbb{R}^d$.
\end{definition}

\noindent The standard Gaussian mechanism adds continuous noise from a Gaussian distribution with magnitude scaled according to \Cref{lem:standard-gaussian-mechanism}.

\begin{definition}[Gaussian Distribution]
\noindent The zero-centered Gaussian distribution (Normal distribution) $\Nd(0, \sigma^2)$ with variance $\sigma^2$ has pdf $\phi(x, \sigma^2) = \frac{1}{2\pi \sigma^2}\exp(-\frac{x^2}{2\sigma^2})$.
We denote the cdf of the Gaussian distribution as $\Phi(X)$.
We refer to $\Phi^{-1}$ as the inverse CDF of the Gaussian distribution.
\end{definition}

\begin{lemma}[{\cite[Theorem~8]{Balle18-AnalyticalGaussian}} The Gaussian Mechanism]
    \label{lem:standard-gaussian-mechanism}
    Let $f:\mathcal{U}^{\mathbb{N}} \rightarrow \mathbb{R}^d$ denote a function with $\ell_2$ sensitivity at most $\Delta_2f$.
    Then the mechanism that outputs $f(X) + Z$, where $Z \sim \Nd(0, \sigma^2 \vec{I}_d)$, satisfies $(\varepsilon, \delta)$-differential privacy if the following inequality holds
    \[
        \Phi \left( \frac{\Delta_2 f}{2\sigma} - \frac{\varepsilon \sigma}{\Delta_2 f} \right) - e^\varepsilon \Phi \left( - \frac{\Delta_2 f}{2\sigma} - \frac{\varepsilon \sigma}{\Delta_2 f} \right) \leq \delta \,.
    \]
\end{lemma}

%% file: content/gshm.tex
\subsection{The Gaussian Sparse Histogram Mechanism}

We consider the problem of privately releasing the histogram $H(\X)$ of a dataset ${\X = (X_1, \cdots, X_n)}$ where each $X_i \in \{0,1\}^{d}$.
The standard techniques for releasing a private histogram are the well-known Laplace mechanism \cite{dwork06calibrating} and Gaussian mechanism~\cite{DworkKMMN06OurDataOurselves, Balle18-AnalyticalGaussian}, which achieve $\epsilon$- and $(\epsilon, \delta)$-DP, respectively.
Although this works well for dense data, it is unsuited for very sparse data where $\|H(\X)\|_0 \ll d$ because adding noise to \emph{all entries} increases both the maximum error as well as the space and time requirements.
Furthermore, the mechanisms are undefined for infinite domains ($d = \infty$).

In the classic sparse histogram setting where $\|X_i\|_0 = 1$ under $(\varepsilon, \delta)$-DP, the preferred technique is the \emph{stability histogram} which combines Laplace noise with a thresholding technique~(see \cite{BalcerVadhan-DP-finite-computers, Korolova09-DP-approx-sparse-hist, BunNS16-simultaneous-learning-multiple-concepts,ALP,googlelibthreshold,Gotz2012,WilsonZLDSG20-SQL-bounded-contribution}). 

In this paper, we consider the \emph{Gaussian Sparse Histogram Mechanism}~(see \cite{googlelibthreshold, Wilkins24-GaussianSparseHistogramMechanism}), which replaces Laplace noise in the stability histogram with Gaussian noise. 
This is often preferred when users can contribute multiple items as the magnitude of noise is scaled by the $\ell_2$ sensitivity instead of the $\ell_1$ sensitivity.
The GSHM adds Gaussian noise to each non-zero counter of $H(\X)$ and removes all counters below a threshold $1 + \tau$.
Find the pseudocode in \cref{alg:uncorrelated-GSHM}.
We discuss the impact of the parameter $\tau$ below.

\begin{algorithm}[h]
    \caption{The Gaussian Sparse Histogram Mechanism~(GSHM)}
    \label{alg:uncorrelated-GSHM}
    \begin{algorithmic}[1]
        \Require Parameters $\sigma$ and $\tau$, histogram $H(\X) \in \mathbb{N}^d$. 
        \State Let $\vec{\tilde{H}} = \{0\}^d$.
        \For{each $i \in [d]$ where $H(\X)_i \neq 0$}
        \State Sample $Z_i \sim \Nd\left(0, \sigma^2\right)$.
        \If{$H(\X)_i + Z_i > 1 + \tau$}
        \State Set $\tilde{H}_i = H(\X)_i + Z_i$.
        \EndIf
        \EndFor
        \State \textbf{Release} $\vec{\tilde{H}}$. 
    \end{algorithmic}
\end{algorithm}

To get $(\epsilon, \delta)$ differential privacy guarantees, we observe that there are two sources of privacy loss that have to be accounted for by the value of $\delta$: 
$\deltaone$ from the Gaussian noise itself and the probability of infinite privacy loss $\deltatwo$, when a zero count is ignored in one dataset, but possibly released in a neighboring one.
These events have infinite privacy loss because they can only occur for one of the datasets.
Therefore, $\deltatwo$ bounds the possibility of outputting a counter that is not present in the neighboring dataset.
\cref{fig:intermediate-histogram} gives some intuition about the role of $\deltaone$ and $\deltatwo$.
Throughout this section, we assume that $\|X_i\|_0 \leq k$ for some known integer $k \ll d$.
Next, we discuss two approaches for obtaining the overall privacy guarantees of the GSHM.

\subsection{The \atd~ Approach}
The following approach appeared in a technical report in the \emph{Google differential privacy library} \cite{googlelibthreshold}. 
Similar techniques have been used elsewhere in the literature (e.g. \cite{WilsonZLDSG20-SQL-bounded-contribution}).
We refer to this technique as \atd. %
We have to account for the privacy loss due to the magnitude of the Gaussian noise.
Therefore, the value of $\deltaone$ is typically found by considering the worst-case effect of a user that only changes non-zero counters in both histograms. 
The value of $\deltaone$ follows from applying \Cref{lem:standard-gaussian-mechanism} (compare also \cref{fig:neighboring-histograms}\textbf{a)}).

The event of infinite privacy loss is captured by $\deltatwo$. 
This is bounded by considering the worst-case scenario of changing $k$ zero-counters to $1$ and the fact that infinite privacy loss occurs exactly when any of these $k$ counters exceed the threshold in the non-zero dataset (compare \cref{fig:neighboring-histograms}\textbf{b)}).

The observation for the \atd~approach is that $\deltaone + \deltatwo$ is a valid upper bound on the overall $\delta$ value, and hence the condition of \Cref{lem:related-add-the-delta} is sufficient to achieve differential privacy. %

{\begin{lemma}[\cite{googlelibthreshold} \atd]
    \label{lem:related-add-the-delta}
    If any pair of neighboring histograms differs in at most $k$ counters 
    then the Gaussian Sparse Histogram Mechanism with parameters $\sigma$ and $\tau$ satisfies $(\varepsilon, \deltaone + \deltatwo)$-differential privacy where
    \begin{align*}
        \deltaone &=  \Phi \left( \frac{\sqrt{k}}{2\sigma} - \frac{\varepsilon \sigma}{\sqrt{k}} \right) - e^\varepsilon \Phi \left( - \frac{\sqrt{k}}{2\sigma} - \frac{\varepsilon \sigma}{\sqrt{k}} \right) \\
        \deltatwo &=  1 - \Phi \left( \frac{\tau}{\sigma} \right)^{k} \,.
    \end{align*}
\end{lemma}
}

\subsection{Exact Analysis by Taking the Max over the Sensitivity Space}

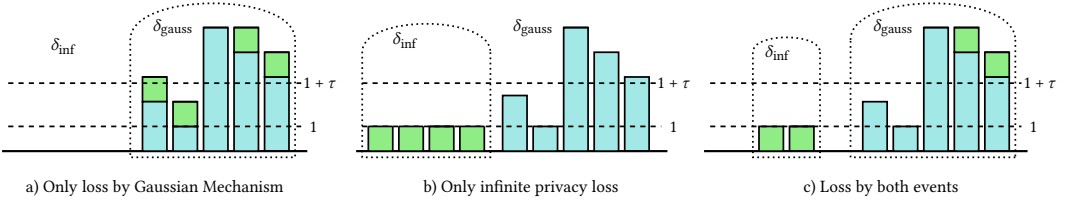
\begin{figure}[t]
    \centering
    \scalebox{0.65}{
        \input{fig/gshn-privacy-events.tikz}
    }
    \caption{
    Examples of different kinds of neighboring datasets for the Gaussian Sparse Histogram Mechanism where a single user can contribute to at most four counters, thus $\|X_i\|_0 \leq 4$.
    These counters are depicted in green.
    \textbf{a)} For the example on the left, the mechanism behaves exactly as running the Gaussian mechanism on a restricted domain. 
    \textbf{b)} In the case in the middle, we only have to bound the probability that one of the green elements together with the additive noise term exceeds the threshold $1 + \tau$. 
    \textbf{c)} The case on the right is the most difficult case for the privacy analysis because the overall $\delta$ value depends on both kinds of changes. 
    }
    \label{fig:neighboring-histograms}
\end{figure}

While the \atd~ approach is sufficient to bound $\delta$ it does not give the tightest possible parameters.
\citet{Wilkins24-GaussianSparseHistogramMechanism} were able to derive an exact value for $\delta$.
Compared to the \atd~approach above, the key insight is that the two worst-case scenarios cannot occur simultaneously for any pair of neighboring histograms.
This means that each counter either contributes to $\deltaone$ \emph{or} $\deltatwo$, but never to both.
In the extreme cases, either all $k$ counters flip from zero to one, from one to zero, or none of them do between neighboring datasets. 
Thus, we only need to consider a single source of privacy loss.
In the other (mixed) cases, we have to consider both sources of privacy loss, but the change is smaller than for the worst-case pair of datasets. 
A small example is depicted in \Cref{fig:neighboring-histograms}.
\citet{Wilkins24-GaussianSparseHistogramMechanism} used this fact to reduce the threshold required to satisfy the given privacy parameters using a tighter analysis.
We now restate their main result: the exact privacy analysis of the GSHM.

\begin{lemma}[\cite{Wilkins24-GaussianSparseHistogramMechanism} Exact Privacy Analysis of the GSHM]
    \label{lem:related-GSHM}
    If any pair of neighboring histograms differs in at most $k$ counters 
    then the Gaussian Sparse Histogram Mechanism with parameters $\sigma$ and $\tau$ satisfies $(\varepsilon, \delta)$-differential privacy where $\gamma(j) = (k - j)\log \Phi \left(\frac{\tau}{\sigma}\right)$ and
    \begin{align*}
        \max \bigg[ &1 - \Phi \left( \frac{\tau}{\sigma} \right)^k, \\
        &\max_{j \in [k]} 1 - \Phi \left( \frac{\tau}{\sigma} \right)^{k - j} + \Phi \left( \frac{\tau}{\sigma} \right)^{k - j} \left[ \Phi\left(\frac{\sqrt{j}}{2\sigma} - \frac{(\varepsilon - \gamma(j))\sigma}{\sqrt{j}}\right) - e^{\varepsilon - \gamma(j)} \Phi\left(-\frac{\sqrt{j}}{2\sigma} - \frac{\left(\varepsilon - \gamma(j)\right)\sigma}{\sqrt{j}}\right) \right], \\
        &\max_{j \in [k]} \Phi\left(\frac{\sqrt{j}}{2\sigma} - \frac{\left(\varepsilon + \gamma(j)\right)\sigma}{\sqrt{j}}\right) - e^{\varepsilon + \gamma(j)} \Phi\left(-\frac{\sqrt{j}}{2\sigma} - \frac{(\varepsilon + \gamma(j))\sigma}{\sqrt{j}}\right)
        \bigg]
        \leq \delta \,.
    \end{align*}
\end{lemma}

\noindent Note that we use a slightly different convention from \citet{Wilkins24-GaussianSparseHistogramMechanism}. 
We use a threshold of $1 + \tau$ rather than $\tau$.
Furthermore, they also consider a more general mechanism that allows for optional aggregation queries and an additional threshold parameter. %
Although our work can easily be adapted to this setting as well, these extensions are not the focus of our main work to simplify presentation.
A discussion of the extensions can be found in \Cref{sec:aggregators}.

\subsection{The Correlated Gaussian Mechanism}
Our work builds on a recent result by \citet{lebeda2024} about using correlated noise to improve the utility of the Gaussian mechanism under the add-remove neighboring relationship.
They show that when answering $d$ counting queries, adding a small amount of correlated noise to all queries can reduce the magnitude of Gaussian noise by almost half.
We restate their main result for $(\varepsilon, \delta)$-differential privacy with a short proof as they use another privacy notion definition.

\begin{lemma}[\cite{lebeda2024} The Correlated Gaussian Mechanism]
    \label{lem:restated-Leb24-for-approx}
    Let $H(\X) \coloneqq \sum_{i = 1}^n X_i$ where $X_i \in \{0,1\}^d$.
    Then the mechanism that outputs $H(\X) + Z_{cor}{\textbf{1}^d} + Z$ satisfies $(\varepsilon, \delta)$-DP.
    Here ${\textbf{1}^d}$ is the $d$-dimensional vector of all ones, $Z_{cor} \sim \Nd(0, \sigma^2/\gamma)$, and $Z \sim \Nd(0, \sigma^2 \vec{I}_d)$ where %
    \[
        \Phi \left( \frac{\sqrt{d + \gamma}}{4\sigma} - \frac{2\varepsilon \sigma}{\sqrt{d + \gamma}} \right) - e^\varepsilon \Phi \left( - \frac{\sqrt{d + \gamma}}{4\sigma} - \frac{2\varepsilon \sigma}{\sqrt{d + \gamma}} \right) \leq \delta \,.
    \] 
\end{lemma}

\begin{proof}
    This follows from combining the inequality for the standard Gaussian mechanism (see \Cref{lem:standard-gaussian-mechanism}) with \cite[Lemma~3.5]{lebeda2024}.
    Furthermore, if we set $\gamma = \sqrt{d}$ as in \cite[Theorem~3.1]{lebeda2024} we minimize the total magnitude of noise.
    Notice that the value of $\sigma$ scales with $\sqrt{d}$ for the standard Gaussian mechanism and it scales with $\frac{1}{2}\sqrt{d + \gamma} = \frac{1}{2}\sqrt{d + \sqrt{d}}$ here.
    We add two noise samples to each query, and the total error for each data point scales with $(\sqrt{d} + 1)/2$.
\end{proof}

Concurrently with the result discussed above, \citet{JosephYu2024-constructions-k-norm-elliptic} considered the setting where user contributions are sparse such that $\|X_i\|_0 \leq k$ for some $k \leq d/2$.
They give an algorithm that adds the optimal amount of correlated Gaussian noise for any $d$ and $k \leq d/2$.
It is natural to ask whether this algorithm can improve error for our setting since they focus on a sparse setting. 
However, their improvement factor over the standard Gaussian mechanism depends on the sparsity. 
Naively applying their technique for the setting of the GSHM where $k \ll d$ yields no practical improvements.
We instead focus on adapting the technique from \cite{lebeda2024}.

%% file: fig/gshn-privacy-events.tikz
\begin{tikzpicture}
	\begin{pgfonlayer}{nodelayer}
		\node [style=none] (0) at (10.5, 0) {};
		\node [style=none] (34) at (3.75, 3) {};
		\node [style=none] (35) at (4.75, 3) {};
		\node [style=none] (36) at (4.75, 2) {};
		\node [style=none] (37) at (3.75, 2) {};
		\node [style=none] (46) at (9.75, 0) {};
		\node [style=none] (47) at (8.75, 0) {};
		\node [style=none] (48) at (8.75, 3) {};
		\node [style=none] (49) at (9.75, 3) {};
		\node [style=none] (158) at (-2, 0) {};
		\node [style=none] (159) at (10.25, 1) {};
		\node [style=none] (160) at (-1.75, 1) {};
		\node [style=none] (161) at (7.25, 0) {};
		\node [style=none] (162) at (6.25, 0) {};
		\node [style=none] (163) at (6.25, 5) {};
		\node [style=none] (164) at (7.25, 5) {};
		\node [style=none] (165) at (8.5, 0) {};
		\node [style=none] (166) at (7.5, 0) {};
		\node [style=none] (167) at (7.5, 4) {};
		\node [style=none] (168) at (8.5, 4) {};
		\node [style=none] (170) at (6, 0) {};
		\node [style=none] (171) at (5, 0) {};
		\node [style=none] (172) at (5, 1) {};
		\node [style=none] (173) at (6, 1) {};
		\node [style=none] (174) at (4.75, 0) {};
		\node [style=none] (175) at (3.75, 0) {};
		\node [style=none] (176) at (3.75, 2) {};
		\node [style=none] (177) at (4.75, 2) {};
		\node [style=none] (178) at (10.25, 2.75) {};
		\node [style=none] (179) at (-1.75, 2.75) {};
		\node [style=none] (180) at (11, 2.75) {$1+ \tau$};
		\node [style=none] (181) at (10.75, 1) {$1$};
		\node [style=none] (182) at (5, 2) {};
		\node [style=none] (183) at (6, 2) {};
		\node [style=none] (184) at (6, 1) {};
		\node [style=none] (185) at (5, 1) {};
		\node [style=none] (186) at (7.5, 5) {};
		\node [style=none] (187) at (8.5, 5) {};
		\node [style=none] (188) at (8.5, 4) {};
		\node [style=none] (189) at (7.5, 4) {};
		\node [style=none] (190) at (8.75, 4) {};
		\node [style=none] (191) at (9.75, 4) {};
		\node [style=none] (192) at (9.75, 3) {};
		\node [style=none] (193) at (8.75, 3) {};
		\node [style=none] (196) at (4.25, 3) {};
		\node [style=none] (197) at (5.5, 2) {};
		\node [style=none] (198) at (8, 5) {};
		\node [style=none] (199) at (9.25, 4) {};
		\node [style=none] (200) at (6.75, 5) {};
		\node [style=none] (201) at (4.25, 1.5) {};
		\node [style=none] (203) at (6.75, 4) {};
		\node [style=none] (204) at (8, 3.5) {};
		\node [style=none] (221) at (6.25, 5) {};
		\node [style=none] (222) at (7.25, 5) {};
		\node [style=none] (223) at (7.5, 5) {};
		\node [style=none] (224) at (8.5, 5) {};
		\node [style=none] (225) at (8.75, 4) {};
		\node [style=none] (226) at (9.75, 4) {};
		\node [style=none] (227) at (3.75, 3) {};
		\node [style=none] (228) at (4.75, 3) {};
		\node [style=none] (229) at (5, 2) {};
		\node [style=none] (230) at (6, 2) {};
		\node [style=none] (232) at (40, 0) {};
		\node [style=none] (234) at (30.25, 1) {};
		\node [style=none] (235) at (31.25, 1) {};
		\node [style=none] (236) at (31.25, 0) {};
		\node [style=none] (237) at (30.25, 0) {};
		\node [style=none] (239) at (39.25, 0) {};
		\node [style=none] (240) at (38.25, 0) {};
		\node [style=none] (241) at (38.25, 3) {};
		\node [style=none] (242) at (39.25, 3) {};
		\node [style=none] (243) at (26.75, 0) {};
		\node [style=none] (244) at (39.75, 1) {};
		\node [style=none] (245) at (27, 1) {};
		\node [style=none] (246) at (36.75, 0) {};
		\node [style=none] (247) at (35.75, 0) {};
		\node [style=none] (248) at (35.75, 5) {};
		\node [style=none] (249) at (36.75, 5) {};
		\node [style=none] (250) at (38, 0) {};
		\node [style=none] (251) at (37, 0) {};
		\node [style=none] (252) at (37, 4) {};
		\node [style=none] (253) at (38, 4) {};
		\node [style=none] (255) at (35.5, 0) {};
		\node [style=none] (256) at (34.5, 0) {};
		\node [style=none] (257) at (34.5, 1) {};
		\node [style=none] (258) at (35.5, 1) {};
		\node [style=none] (259) at (34.25, 0) {};
		\node [style=none] (260) at (33.25, 0) {};
		\node [style=none] (261) at (33.25, 2) {};
		\node [style=none] (262) at (34.25, 2) {};
		\node [style=none] (263) at (39.75, 2.75) {};
		\node [style=none] (264) at (27, 2.75) {};
		\node [style=none] (265) at (40.5, 2.75) {$1 + \tau$};
		\node [style=none] (266) at (40.25, 1) {$1$};
		\node [style=none] (267) at (29, 1) {};
		\node [style=none] (268) at (30, 1) {};
		\node [style=none] (269) at (30, 0) {};
		\node [style=none] (270) at (29, 0) {};
		\node [style=none] (271) at (37, 5) {};
		\node [style=none] (272) at (38, 5) {};
		\node [style=none] (273) at (38, 4) {};
		\node [style=none] (274) at (37, 4) {};
		\node [style=none] (275) at (38.25, 4) {};
		\node [style=none] (276) at (39.25, 4) {};
		\node [style=none] (277) at (39.25, 3) {};
		\node [style=none] (278) at (38.25, 3) {};
		\node [style=none] (279) at (30.75, 1) {};
		\node [style=none] (280) at (35, 1) {};
		\node [style=none] (281) at (37.5, 5) {};
		\node [style=none] (282) at (38.75, 4) {};
		\node [style=none] (283) at (36.25, 5) {};
		\node [style=none] (286) at (36.25, 4) {};
		\node [style=none] (287) at (37.5, 3.5) {};
		\node [style=none] (299) at (35.75, 5) {};
		\node [style=none] (300) at (36.75, 5) {};
		\node [style=none] (301) at (37, 5) {};
		\node [style=none] (302) at (38, 5) {};
		\node [style=none] (303) at (38.25, 4) {};
		\node [style=none] (304) at (39.25, 4) {};
		\node [style=none] (305) at (30.25, 1) {};
		\node [style=none] (306) at (31.25, 1) {};
		\node [style=none] (310) at (25.25, 0) {};
		\node [style=none] (313) at (19.5, 2.25) {};
		\node [style=none] (316) at (22.25, 0) {};
		\node [style=none] (317) at (24.5, 0) {};
		\node [style=none] (318) at (23.5, 0) {};
		\node [style=none] (319) at (23.5, 3) {};
		\node [style=none] (320) at (24.5, 3) {};
		\node [style=none] (321) at (12.5, 0) {};
		\node [style=none] (322) at (25, 1) {};
		\node [style=none] (323) at (12.75, 1) {};
		\node [style=none] (324) at (22, 0) {};
		\node [style=none] (325) at (21, 0) {};
		\node [style=none] (326) at (21, 5) {};
		\node [style=none] (327) at (22, 5) {};
		\node [style=none] (328) at (23.25, 0) {};
		\node [style=none] (329) at (22.25, 0) {};
		\node [style=none] (330) at (22.25, 4) {};
		\node [style=none] (331) at (23.25, 4) {};
		\node [style=none] (332) at (21.75, 0) {};
		\node [style=none] (333) at (20.75, 0) {};
		\node [style=none] (334) at (19.75, 0) {};
		\node [style=none] (335) at (19.75, 1) {};
		\node [style=none] (336) at (20.75, 1) {};
		\node [style=none] (337) at (19.5, 0) {};
		\node [style=none] (338) at (18.5, 0) {};
		\node [style=none] (339) at (18.5, 2.25) {};
		\node [style=none] (340) at (19.5, 2.25) {};
		\node [style=none] (341) at (25, 2.75) {};
		\node [style=none] (342) at (12.75, 2.75) {};
		\node [style=none] (343) at (25.5, 2.75) {$1 + \tau$};
		\node [style=none] (344) at (25.5, 1) {$1$};
		\node [style=none] (352) at (22.25, 4) {};
		\node [style=none] (353) at (15.5, 1) {};
		\node [style=none] (354) at (16.5, 1) {};
		\node [style=none] (355) at (16.5, 0) {};
		\node [style=none] (356) at (15.5, 0) {};
		\node [style=none] (357) at (19, 2.25) {};
		\node [style=none] (358) at (20.25, 1) {};
		\node [style=none] (360) at (16, 1) {};
		\node [style=none] (361) at (21.5, 5) {};
		\node [style=none] (364) at (21.5, 4) {};
		\node [style=none] (377) at (21, 5) {};
		\node [style=none] (378) at (22, 5) {};
		\node [style=none] (381) at (15.5, 1) {};
		\node [style=none] (382) at (16.5, 1) {};
		\node [style=none] (384) at (19.5, 2.25) {};
		\node [style=none] (388) at (29.5, 1) {};
		\node [style=none] (392) at (28.75, 4) {};
		\node [style=none] (393) at (31.5, 4) {};
		\node [style=none] (394) at (31.5, -0.25) {};
		\node [style=none] (395) at (28.75, -0.25) {};
		\node [style=none] (397) at (3.25, 4.5) {};
		\node [style=none] (398) at (10, 4.5) {};
		\node [style=none] (399) at (10, -0.25) {};
		\node [style=none] (400) at (3.25, -0.25) {};
		\node [style=none] (401) at (16.75, 1) {};
		\node [style=none] (402) at (17.75, 1) {};
		\node [style=none] (403) at (17.75, 0) {};
		\node [style=none] (404) at (16.75, 0) {};
		\node [style=none] (405) at (17.25, 1) {};
		\node [style=none] (409) at (16.75, 1) {};
		\node [style=none] (410) at (17.75, 1) {};
		\node [style=none] (411) at (14.25, 1) {};
		\node [style=none] (412) at (15.25, 1) {};
		\node [style=none] (413) at (15.25, 0) {};
		\node [style=none] (414) at (14.25, 0) {};
		\node [style=none] (415) at (14.75, 1) {};
		\node [style=none] (419) at (14.25, 1) {};
		\node [style=none] (420) at (13, 1) {};
		\node [style=none] (421) at (14, 1) {};
		\node [style=none] (422) at (14, 0) {};
		\node [style=none] (423) at (13, 0) {};
		\node [style=none] (424) at (13.5, 1) {};
		\node [style=none] (428) at (13, 1) {};
		\node [style=none] (429) at (12.75, 4) {};
		\node [style=none] (430) at (18, 4) {};
		\node [style=none] (431) at (18, -0.25) {};
		\node [style=none] (432) at (12.75, -0.25) {};
		\node [style=none] (437) at (5, 5) {$\deltaone$};
		\node [style=none] (439) at (32.75, 4.5) {};
		\node [style=none] (440) at (39.5, 4.5) {};
		\node [style=none] (441) at (39.5, -0.25) {};
		\node [style=none] (442) at (32.75, -0.25) {};
		\node [style=none] (445) at (33.75, 2) {};
		\node [style=none] (447) at (24, 3) {};
		\node [style=none] (450) at (22.75, 4) {};
		\node [style=none] (455) at (0.5, 4.25) {$\deltatwo$};
		\node [style=none] (456) at (29.75, 4) {$\deltatwo$};
		\node [style=none] (457) at (34.5, 5) {$\deltaone$};
		\node [style=none] (458) at (19.75, 5) {$\deltaone$};
		\node [style=none] (459) at (14.5, 4.5) {$\deltatwo$};
		\node [style=none] (460) at (4.25, -1.5) {a) Only loss by Gaussian Mechanism};
		\node [style=none] (461) at (34, -1.5) {c) Loss by both events};
		\node [style=none] (462) at (19.25, -1.5) {b) Only infinite privacy loss};
	\end{pgfonlayer}
	\begin{pgfonlayer}{edgelayer}
		\draw [style=FILLGREEN] (37.center)
			 to (34.center)
			 to (35.center)
			 to (36.center)
			 to cycle;
		\draw [style=FILLBLUE] (48.center)
			 to (47.center)
			 to (46.center)
			 to (49.center)
			 to cycle;
		\draw [style=thick] (0.center) to (158.center);
		\draw [style=FILLBLUE] (162.center)
			 to (161.center)
			 to (164.center)
			 to (163.center)
			 to cycle;
		\draw [style=FILLBLUE] (166.center)
			 to (165.center)
			 to (168.center)
			 to (167.center)
			 to cycle;
		\draw [style=FILLBLUE] (171.center)
			 to (170.center)
			 to (173.center)
			 to (172.center)
			 to cycle;
		\draw [style=FILLBLUE] (175.center)
			 to (174.center)
			 to (177.center)
			 to (176.center)
			 to cycle;
		\draw [style=EDGEDASHED] (179.center) to (178.center);
		\draw [style=FILLGREEN] (185.center)
			 to (182.center)
			 to (183.center)
			 to (184.center)
			 to cycle;
		\draw [style=FILLGREEN] (189.center)
			 to (186.center)
			 to (187.center)
			 to (188.center)
			 to cycle;
		\draw [style=FILLGREEN] (192.center)
			 to (193.center)
			 to (190.center)
			 to (191.center)
			 to cycle;
		\draw [style=EDGEDASHED] (160.center) to (159.center);
		\draw [style=FILLBLUE] (221.center) to (222.center);
		\draw [style=FILLBLUE] (223.center) to (224.center);
		\draw [style=FILLBLUE] (225.center) to (226.center);
		\draw [style=FILLBLUE] (227.center) to (228.center);
		\draw [style=FILLBLUE] (229.center) to (230.center);
		\draw [style=FILLGREEN] (237.center)
			 to (234.center)
			 to (235.center)
			 to (236.center)
			 to cycle;
		\draw [style=FILLBLUE] (241.center)
			 to (240.center)
			 to (239.center)
			 to (242.center)
			 to cycle;
		\draw [style=thick] (232.center) to (243.center);
		\draw [style=FILLBLUE] (247.center)
			 to (246.center)
			 to (249.center)
			 to (248.center)
			 to cycle;
		\draw [style=FILLBLUE] (251.center)
			 to (250.center)
			 to (253.center)
			 to (252.center)
			 to cycle;
		\draw [style=FILLBLUE] (256.center)
			 to (255.center)
			 to (258.center)
			 to (257.center)
			 to cycle;
		\draw [style=FILLBLUE] (260.center)
			 to (259.center)
			 to (262.center)
			 to (261.center)
			 to cycle;
		\draw [style=EDGEDASHED] (264.center) to (263.center);
		\draw [style=FILLGREEN] (270.center)
			 to [in=270, out=90] (267.center)
			 to (268.center)
			 to (269.center)
			 to cycle;
		\draw [style=FILLGREEN] (274.center)
			 to (271.center)
			 to (272.center)
			 to (273.center)
			 to cycle;
		\draw [style=FILLGREEN] (277.center)
			 to (278.center)
			 to (275.center)
			 to (276.center)
			 to cycle;
		\draw [style=EDGEDASHED] (245.center) to (244.center);
		\draw [style=FILLBLUE] (299.center) to (300.center);
		\draw [style=FILLBLUE] (301.center) to (302.center);
		\draw [style=FILLBLUE] (303.center) to (304.center);
		\draw [style=FILLBLUE] (305.center) to (306.center);
		\draw [style=FILLBLUE] (319.center)
			 to (318.center)
			 to (317.center)
			 to (320.center)
			 to cycle;
		\draw [style=thick] (310.center) to (321.center);
		\draw [style=FILLBLUE] (325.center)
			 to (324.center)
			 to (327.center)
			 to (326.center)
			 to cycle;
		\draw [style=FILLBLUE] (329.center)
			 to (328.center)
			 to (331.center)
			 to (330.center)
			 to cycle;
		\draw [style=FILLBLUE] (334.center)
			 to (333.center)
			 to (336.center)
			 to (335.center)
			 to cycle;
		\draw [style=FILLBLUE] (338.center)
			 to (337.center)
			 to (340.center)
			 to (339.center)
			 to cycle;
		\draw [style=EDGEDASHED] (342.center) to (341.center);
		\draw [style=FILLGREEN] (355.center)
			 to (356.center)
			 to (353.center)
			 to (354.center)
			 to cycle;
		\draw [style=EDGEDASHED] (323.center) to (322.center);
		\draw [style=FILLBLUE] (377.center) to (378.center);
		\draw [style=FILLBLUE] (381.center) to (382.center);
		\draw [style=black dotted] (393.center)
			 to (394.center)
			 to (395.center)
			 to (392.center)
			 to [bend left=90, looseness=0.75] cycle;
		\draw [style=black dotted] (398.center)
			 to (399.center)
			 to [bend right=360, looseness=0.75] (400.center)
			 to (397.center)
			 to [bend left=90, looseness=0.75] cycle;
		\draw [style=FILLGREEN] (402.center)
			 to (403.center)
			 to (404.center)
			 to (401.center);
		\draw [style=FILLGREEN] (401.center) to (402.center);
		\draw [style=FILLGREEN] (412.center)
			 to (413.center)
			 to (414.center)
			 to (411.center);
		\draw [style=FILLGREEN] (411.center) to (412.center);
		\draw [style=FILLGREEN] (421.center)
			 to (422.center)
			 to (423.center)
			 to (420.center);
		\draw [style=FILLGREEN] (420.center) to (421.center);
		\draw [style=black dotted] (430.center)
			 to (431.center)
			 to (432.center)
			 to (429.center)
			 to [bend left=90, looseness=0.75] cycle;
		\draw [style=black dotted] (440.center) to (441.center);
		\draw [style=black dotted, bend right=360, looseness=0.75] (441.center) to (442.center);
		\draw [style=black dotted] (442.center) to (439.center);
		\draw [style=black dotted, bend left=90, looseness=0.75] (439.center) to (440.center);
	\end{pgfonlayer}
\end{tikzpicture}

%% file: content/results-intro.tex
\section{Algorithmic Framework}
\label{sec:algorithm}

We are now ready to introduce our main contribution, a variant of the Gaussian Sparse Histogram Mechanism using correlated noise. %
Throughout this section, we assume that all histograms are \emph{$k$-sparse monotonic} as defined below.
Intuitively, we define monotonicity on a histogram in a way that captures the setting where the counts are either all increasing or all decreasing.

\begin{definition}[{$k$-sparse monotonic histogram}] \label{def:k-sparse-histogram}
    We assume that the input histogram is $k$-sparse. 
    That is, for any %
    dataset $\X$ we have $\|H(\X)\|_0 \leq k$. %
    Furthermore, the sensitivity space of $H$ is $\{0,1\}^d \cup \{0,-1\}^d$.
    That is, the difference between counters of neighboring histograms are either non-decreasing or non-increasing.
\end{definition}

Observe that due to the monotonicity constraint, we either have that the supports $U$ and $U'$ for two neighboring histograms satisfies $U\subseteq U'$ or $U' \subseteq U$.
We use this property in the privacy proofs later.
The histograms are also monotonic in \cite{Wilkins24-GaussianSparseHistogramMechanism}, but they do not require $k$-sparsity.
We provide a mechanism for a setting where the histograms are not $k$-sparse in \Cref{sec:top-k}.

One example of $k$-sparse monotonic histograms is Misra-Gries sketches.
Merging Misra-Gries sketches is common in practical applications.
The sensitivity space of merged Misra-Gries sketches of size $k$ exactly matches \Cref{def:k-sparse-histogram}~(See \citet{LebedaTetek23-DPMG}).
The exact procedure used to generate the histogram is not important for this section. 
Our algorithm is differentially private as long as the structure between histograms holds for all pairs of neighboring datasets.

Notice that \Cref{def:k-sparse-histogram} implies that neighboring histograms differ in at most $k$ counters.
As such, we can release the histogram %
using the standard Gaussian Sparse Histogram Mechanism. %
\citet{Wilkins24-GaussianSparseHistogramMechanism} already uses the the fact that counters are either non-decreasing or non-increasing in their analysis.
We intend to further take advantage of the monotonicity by adding a small amount of correlated noise to all non-zero counters.
This allows us to reduce the total magnitude of noise similar to \cite{lebeda2024}.
The reduced magnitude of noise in turn allows us to reduce the threshold required for privacy.
The pseudocode for our mechanism is in \Cref{alg:main}.

\begin{algorithm}[h]
    \caption{Correlated Stability Histogram~(\algoname)}\label{alg:main}
    \begin{algorithmic}[1]
        \Require Parameters $k$, $\sigma$ and $\tau$, histogram $H(\X) \in \mathbb{N}^d$ where $\|H(\X)\|_0 \leq k$. 
        \State Let $\vec{\tilde{H}} = \{0\}^d$.
        \State Sample $Z_{\text{corr}} \sim \Nd\left(0, \sigma^2/\sqrt{k}\right)$.
        \For{each $i \in [d]$ where $H(\X)_i \neq 0$}
        \State Sample $Z_i \sim \Nd\left(0, \sigma^2\right)$.
        \If{$H(\X)_i + Z_i + Z_{\text{corr}} > 1 + \tau$}\label{line:main-alg-threshold}
        \State Set $\tilde{H}_i = H(\X)_i + Z_i + Z_{\text{corr}}$.
        \EndIf
        \EndFor
        \State \textbf{Release} $\tilde{H}$. 
    \end{algorithmic}
\end{algorithm}

Next, we prove the algorithm's privacy guarantees. 
We first give privacy guarantees in a (relatively) simple closed-form which is similar to the \atd~ approach. %
Later, we give tighter bounds using a more complicated analysis similar to \Cref{lem:related-GSHM}.
Unfortunately, we cannot reuse the proofs from previous work in either case because they rely on the fact that all noisy counters are independent.
That is clearly not the case for our mechanism because the value of $Z_{cor}$ is added to all entries.
Instead, we use different techniques to give similar results, starting with the \atd~ approach.
The following lemma gives a general bound the event that one of $j$ correlated noisy terms exceeds a threshold $\tau$.
We use this in the proof for both approaches later.

\begin{lemma}[Upper bound for Correlated Noise]\label{claim:upperbounddeltainf}
Let $Z_{\text{Corr}} \sim \Nd(0, \sigma^2/\sqrt{k})$ be a single sample for a real $k > 0$ and together with $j$ additional samples $Z_1, \dots, Z_j \sim \Nd(0, \sigma^2)$.
Then for any $\tau > 0$, we have
\[
\Pr[\exists i \in [j]: Z_{\text{corr}} + Z_i > \tau] \leq 1 - \Phi\left(\dfrac{\tau}{\sigma\left(k^{-1/4}+1\right)}\right)^{j+1}\,.
\]
\end{lemma}
\begin{proof}
We first give a bound for $Z_{\text{corr}}$ and the $Z_i$'s independently, which is sufficient for us to bound the probability of this event.
First, observe that the $k$ uncorrelated terms $Z_i \sim \Nd(0, \sigma^2)$ are independent, and we are interested in bounding the maximum of them and hence: 
\begin{align*}
        \Pr\left[\max\limits_{Z_1, \cdots, Z_j} Z_i \leq \frac{\tau}{1+k^{-1/4}}\right]  = \Pr\left[\Nd(0, \sigma^2) \leq \frac{\tau}{1+k^{-1/4}}\right]^j
        =\Phi\left(\frac{\tau}{\sigma\left(1+k^{-1/4}\right)}\right)^j \,.
\end{align*}

Similarly, we have for $Z_{\text{corr}} \sim N(0, \sigma^2/\sqrt{k})$ that
\begin{align*}
\Pr\left[Z_{\text{corr}} \leq \frac{\tau k^{-1/4}}{1+ k ^{-1/4}}\right] &=  \Phi\left(\frac{\tau k^{-1/4}}{\sigma \left(1+ k^{-1/4}\right)\cdot k^{-1/4}}\right) = \Phi\left(\frac{\tau}{\sigma \left(1+k^{-1/4}\right)}\right) \,.
\end{align*}
Notice that if both $Z_i \leq \tau/(1+k^{-1/4})$ and $Z_{\text{corr}} \leq \tau k^{-1/4}/(1+k^{-1/4})$ holds then $Z_{\text{corr}} + Z_i \leq \tau$.
As such, we can prove that the lemma holds since 
\begin{align}
    \Pr[Z_{\text{corr}} + \max\limits_{Z_1, \cdots, Z_j} Z_i  > \tau]
    &\leq \Pr\left[Z_{\text{corr}} > \dfrac{\tau k^{-1/4}}{1+k^{-1/4} } \vee \max\limits_{Z_1, \cdots, Z_j} Z_i > \dfrac{\tau}{1+k^{-1/4}} \right]\label{eq:union-bound} \\
    &= 1- \Pr\left[Z_{\text{corr}} \leq \frac{\tau k^{-1/4}}{1+k^{-1/4}}\right]\cdot \Pr\left[\max\limits_{Z_1,\cdots,Z_j} Z_i \leq \frac{\tau}{1+k^{-1/4}}\right]\label{eq:independence}\\
    &= 1 - \Phi\left(\dfrac{\tau}{\sigma\left(1 + k^{-1/4}\right)}\right)^{j+1} \nonumber
\end{align}

where step (\ref{eq:union-bound}) holds by a union bound and step (\ref{eq:independence}) holds because the random variables $Z_{\text{Corr}}$ and $\tilde{Z} = \max_{Z_1, \dots, Z_j} Z_i$ are independent.

\end{proof}

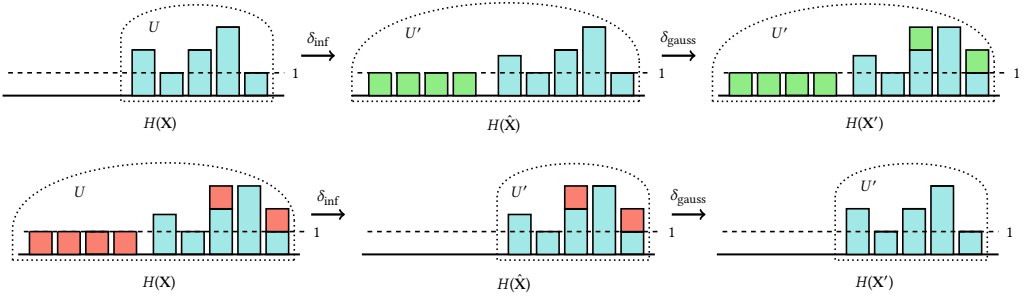
\begin{figure}
    \centering
    \scalebox{0.60}{
        \input{fig/intermediate-dataset.tikz}
    }
    \scalebox{0.60}{
        \input{fig/intermediate-dataset-remove.tikz}
    }
    \caption{The separation technique of the $\deltaone$ and $\deltatwo$ used in \cref{thm:algorithm-privacy-add-the-deltas,th:max-the-delta}.
    The idea is to construct an intermediate histogram $H(\hat{\X})$ with the same support $U'$ as $H(\X')$ but only reflect the changes that can cause infinite privacy loss between $H(\X)$ and $H(\X')$.
    }
    \label{fig:intermediate-histogram}
\end{figure}

\subsection{The \atd~Analysis}

The following \Cref{thm:algorithm-privacy-add-the-deltas} proves privacy guarantees of our mechanism using a similar technique than proposed by \citet{googlelibthreshold} and \citet{WilsonZLDSG20-SQL-bounded-contribution} which is known as \atd~.
The total value for $\delta$ is split between $\deltaone$ and $\deltatwo$
which accounts for the two types of privacy loss that are relevant to the mechanism.
However, like with the GSHM, these values are found by considering worst-case pairs of neighboring histograms for each case.
A pair of neighboring histograms cannot be worst-case for both values as seen in \Cref{fig:neighboring-histograms}. 

\begin{theorem}[\atd~ technique]
    \label{thm:algorithm-privacy-add-the-deltas}
    Algorithm~\ref{alg:main} satisfies $(\varepsilon, \deltaone + \deltatwo)$-differential privacy for $k$-sparse monotonic histograms where
    \begin{align*}
        \deltaone &= \Phi \left( \frac{\sqrt{k + \sqrt{k}}}{4\sigma} - \frac{2\varepsilon \sigma}{\sqrt{k + \sqrt{k}}} \right) - e^\varepsilon \Phi \left( - \frac{\sqrt{k + \sqrt{k}}}{4\sigma} - \frac{2\varepsilon \sigma}{\sqrt{k + \sqrt{k}}} \right) \\
        \deltatwo &= 1 - \Phi \left(\frac{\tau}{\sigma \left(1 + k^{-1/4}\right)} \right)^{k + 1} \,.
    \end{align*}
\end{theorem}

\begin{proof}
    By \Cref{def:differential-privacy}, the lemma holds if for any pair of $k$-sparse monotonic histograms $H(\X)$ and $H(\X')$ and all sets of outputs $Y$ we have
    \[
        \Pr[\algoname(H(\X)) \in Y] \leq e^\varepsilon \Pr[\algoname(H(\X')) \in Y] + \deltaone + \deltatwo \,.
    \]

    We prove that the inequality above holds by introducing a third histogram.
    This new histogram is constructed to be between $H(\X)$ and $H(\X')$.
    We first state the desired properties of this histogram and then show that such a histogram must exist for all neighboring histograms.

    Assume for now that there exists a histogram $H(\hat{\X}) \in \mathbb{N}^d$ where the following two inequalities hold for any set of outputs $Y$: %
    \begin{equation}
        \label{eq:histograms-delta-1}
        \Pr[\algoname(H(\hat{\X})) \in Y] \leq e^\varepsilon \Pr[\algoname(H(\X')) \in Y] + \deltaone \,,
    \end{equation}
    \begin{equation}
        \label{eq:histograms-delta-2}
        \Pr[\algoname(H(\X)) \in Y] \leq \Pr[\algoname(H(\hat{\X})) \in Y] + \deltatwo \,.
    \end{equation}
    Then the inequality we need for the lemma follows immediately since 
    \begin{align*}
        \Pr[\algoname(H(\X)) \in Y] &\leq \Pr[\algoname(H(\hat{\X})) \in Y] + \deltatwo \\ 
        &\leq e^\varepsilon \Pr[\algoname(H(\X')) \in Y] + \deltaone  + \deltatwo \,.
    \end{align*}

    Next, we show how to construct $H(\hat{\X})$ from $H(\X)$ and $H(\X')$, and finally we show that each of two inequalities holds using this construction.
    Let $U = \{i \in [d] : H(\X)_i \neq 0 \}$ denote the support of $H(\X)$ and define similarly $U'$ and $\hat{U}$.
    We construct $H(\hat{\X})$ such that it has the same support as $H(\X')$, that is $U' = \hat{U}$.
    For all $i \in (U \cap U')$ we set $H(\hat{\X})_i = H({\X})_i$ and for all $i \notin (U \cap U')$ we set $H(\hat{\X})_i = H(\X')_i$. 
    In other words, we construct $H(\hat{\X})$ such that 
    (1) $H(\X')$ and $H(\hat{\X})$ only differ in entries that are in both $U$ and $U'$
    (2) $H({\X})$ and $H(\hat{\X})$ only differ in entries that are in only one of $U$ and $U'$.
    This allows us to analyze each case separately to derive our values for $\deltaone$ and $\deltatwo$.
    An example of this construction is shown in \Cref{fig:intermediate-histogram}.

    We start with the case of $\deltaone$ using $H(\X')$ and $H(\hat{\X})$.
    Let ${\algoname'}$ denote a new mechanism equivalent to \Cref{alg:main} except that the condition on line \ref{line:main-alg-threshold} is removed. 
    Notice that with $H(\X')$ or $H(\hat{\X})$ as inputs ${\algoname'}$ is equivalent to the Correlated Gaussian Mechanism restricted to $U'$.
    Since $\vert U' \vert \leq k$ we have by \Cref{lem:restated-Leb24-for-approx} that
    \[
    \Pr[{\algoname}'(H(\hat{\X})) \in Y] \leq e^\varepsilon \Pr[{\algoname}'(H(\X')) \in Y] + \deltaone \,.
    \]
    Notice that if we post-process the output of  ${\algoname}'$ by removing entries below $1 + \tau$, then the output distribution is equivalent to $\algoname$.
    \Cref{eq:histograms-delta-1} therefore holds because post-proccessing does not affect differential privacy guarantees (see \Cref{def:differential-privacy}).

    The histograms $H(\X)$ and $H(\hat{\X})$ only differ in entries that all has a count of $1$ in one of the histograms while they all have a count of $0$ in the other.
    $\deltatwo$ accounts for the event where any such counter exceeds the threshold because the distributions are identical for the shared support.
    The probability of this event is increasing in the number of different elements between $H(\X)$ and $H(\hat{\X})$ and therefore, the worst case happens for neighboring datasets such that $H(\X) = \vec{1}^k$ and $H(\X') = \vec{0}^k$.
    Note that this bound also hold when zero counters in $H(\X)$ are non-zero in $H(\hat{\X})$.
    We focus on one direction because the proof is almost identical for the symmetric case.
    We bound the probability of outputting any entries from $H(\X)$, that is $\tilde{H}_i := 1 + Z_{\text{corr}} + Z_i > 1 + \tau$ for at least one $i \in (U \setminus U')$.
    The bound for $\deltatwo$ follows from setting $j = k$ in \Cref{claim:upperbounddeltainf}.
\end{proof}

\noindent In the \cref{lem:computing-threshold-add-the-delta} below, we show how to directly compute the required threshold based on the result in \Cref{thm:algorithm-privacy-add-the-deltas}.

\begin{lemma}[Computing the threshold $\tau$]
\label{lem:computing-threshold-add-the-delta}
For a fixed privacy budget $\epsilon, \delta$ and parameters $k$ and $\sigma$, the \atd~ technique requires $\tau \geq \Phi^{-1}\left(\sqrt[k+1]{1-\delta-\delta_{\text{gauss}}}\right) \cdot \left(1+k^{-1/4}\right) \sigma$ where $\Phi^{-1}$ is the inverse CDF of the normal distribution and $\deltaone$ is defined as in \Cref{thm:algorithm-privacy-add-the-deltas}.
\end{lemma}
\begin{proof}
Observe $\delta - \delta_{\text{gauss}} \geq  \delta_{\text{inf}} =  1- \Phi(\frac{\tau}{\sigma (1+k^{-1/4})})^{k+1}$ and hence $    {\frac{\tau}{\sigma (1+k^{-1/4})} \geq \Phi^{-1} \left(\sqrt[k+1]{1-\delta - \deltaone}\right)}$
Multiplying by $\sigma \left(1+k^{-1/4}\right)$ gives the desired result.
\end{proof}

\subsection{Tighter Analysis} %
Next, we carry out a more careful analysis that considers all elements from the sensitivity space similar to~\citet{Wilkins24-GaussianSparseHistogramMechanism}.
As discussed above, we cannot directly translate the analysis by \citeauthor{Wilkins24-GaussianSparseHistogramMechanism} because they rely on independence between each entry.
The proof of \Cref{thm:privacy-tighter-analysis} is in \Cref{app:proof-tighter-analysis}. %
Here we give a short intuition behind the theorem.

\begin{restatable}[Tighter Analysis]{theorem}{tighteranalysis}\label{th:max-the-delta}
    \label{thm:privacy-tighter-analysis}
    Let $\gamma(j) = \min\left(\sqrt{j}, \dfrac{1}{2}\sqrt{j + \sqrt{k}}\right)$, $\psi(m) =  \Phi \left( \frac{\tau}{(1 + k^{-1/4})\sigma} \right)^{m+1}$, and $\hat{\varepsilon}(j) = \varepsilon + \ln\left(\psi{(k-j)}\right)$.
    Then \cref{alg:main} with parameters $k$, $\sigma$, and $\tau$ satisfies $(\varepsilon, \delta)$-differential privacy for $k$-sparse monotonic histograms, where 
    \begin{align*}
        \max \bigg[ 1 - \psi(k),~ 
        &\Phi \left( \frac{\sqrt{k + \sqrt{k}}}{4\sigma} - \frac{2\varepsilon \sigma}{\sqrt{k + \sqrt{k}}} \right) - e^\varepsilon \Phi \left( - \frac{\sqrt{k + \sqrt{k}}}{4\sigma} - \frac{2\varepsilon \sigma}{\sqrt{k + \sqrt{k}}} \right) , \\
        &\max_{j \in [k - 1]} 
        1 - \psi({k-j}) + \Phi \left( \frac{\gamma(j)}{2\sigma} - \frac{\varepsilon \sigma}{\gamma(j)} \right) - e^\varepsilon \Phi \left( - \frac{\gamma(j)}{2\sigma} - \frac{\varepsilon \sigma}{\gamma(j)} \right) \,, \\
        &\max_{j \in [k - 1]} \Phi \left( \frac{\gamma(j)}{2\sigma} - \frac{\hat{\varepsilon}(j) \sigma}{\gamma(j)} \right) - e^{\hat{\varepsilon}(j)} \Phi \left( - \frac{\gamma(j)}{2\sigma} - \frac{\hat{\varepsilon}(j) \sigma}{\gamma(j)} \right) \bigg] \leq \delta \,.
    \end{align*}
\end{restatable}

The result relies on a case by case analysis where each term in the maximum corresponds to a specific difference between neighboring histograms.
The first two terms covers cases when we only have to consider either the infinite privacy loss event or the Gaussian noise, respectively. 
The remaining terms cover the differences when we have to account for both Gaussian noise and the threshold similar to case c) in \Cref{fig:neighboring-histograms}.
The first of the internal maximum covers the cases where $U \supset U'$ and $\vert U' \vert \leq j$, while the second covers the cases where $U \subset U'$ and $\vert U \vert \leq j$.
For each case in our analysis, we split up the impact from Gaussian noise and the threshold. 
Together, these cases cover all elements in the sensitivity space for $k$-sparse monotonic histograms.

%% file: fig/intermediate-dataset.tikz
\begin{tikzpicture}
	\begin{pgfonlayer}{nodelayer}
		\node [style=none] (0) at (10.75, 0) {};
		\node [style=none] (3) at (5, 2) {};
		\node [style=none] (4) at (4, 2) {};
		\node [style=none] (5) at (10, 0) {};
		\node [style=none] (6) at (9, 0) {};
		\node [style=none] (7) at (9, 1) {};
		\node [style=none] (8) at (10, 1) {};
		\node [style=none] (9) at (-1.75, 0) {};
		\node [style=none] (10) at (10.75, 1) {};
		\node [style=none] (11) at (-1.5, 1) {};
		\node [style=none] (12) at (7.5, 0) {};
		\node [style=none] (13) at (6.5, 0) {};
		\node [style=none] (14) at (6.5, 2) {};
		\node [style=none] (15) at (7.5, 2) {};
		\node [style=none] (16) at (8.75, 0) {};
		\node [style=none] (17) at (7.75, 0) {};
		\node [style=none] (18) at (7.75, 3) {};
		\node [style=none] (19) at (8.75, 3) {};
		\node [style=none] (20) at (6.25, 0) {};
		\node [style=none] (21) at (5.25, 0) {};
		\node [style=none] (22) at (5.25, 1) {};
		\node [style=none] (23) at (6.25, 1) {};
		\node [style=none] (24) at (5, 0) {};
		\node [style=none] (25) at (4, 0) {};
		\node [style=none] (26) at (4, 2) {};
		\node [style=none] (27) at (5, 2) {};
		\node [style=none] (31) at (11.25, 1) {$1$};
		\node [style=none] (34) at (6.25, 1) {};
		\node [style=none] (35) at (5.25, 1) {};
		\node [style=none] (38) at (8.75, 3) {};
		\node [style=none] (39) at (7.75, 3) {};
		\node [style=none] (42) at (10, 1) {};
		\node [style=none] (43) at (9, 1) {};
		\node [style=none] (48) at (7, 2) {};
		\node [style=none] (54) at (6.5, 2) {};
		\node [style=none] (55) at (7.5, 2) {};
		\node [style=none] (64) at (26.75, 0) {};
		\node [style=none] (69) at (26.25, 0) {};
		\node [style=none] (70) at (25.25, 0) {};
		\node [style=none] (71) at (25.25, 1) {};
		\node [style=none] (72) at (26.25, 1) {};
		\node [style=none] (73) at (13.75, 0) {};
		\node [style=none] (74) at (27, 1) {};
		\node [style=none] (75) at (14, 1) {};
		\node [style=none] (76) at (23.75, 0) {};
		\node [style=none] (77) at (22.75, 0) {};
		\node [style=none] (78) at (22.75, 2) {};
		\node [style=none] (79) at (23.75, 2) {};
		\node [style=none] (80) at (25, 0) {};
		\node [style=none] (81) at (24, 0) {};
		\node [style=none] (82) at (24, 3) {};
		\node [style=none] (83) at (25, 3) {};
		\node [style=none] (84) at (22.5, 0) {};
		\node [style=none] (85) at (21.5, 0) {};
		\node [style=none] (86) at (21.5, 1) {};
		\node [style=none] (87) at (22.5, 1) {};
		\node [style=none] (88) at (21.25, 0) {};
		\node [style=none] (89) at (20.25, 0) {};
		\node [style=none] (90) at (20.25, 1.75) {};
		\node [style=none] (91) at (21.25, 1.75) {};
		\node [style=none] (95) at (27.5, 1) {$1$};
		\node [style=none] (102) at (25, 3) {};
		\node [style=none] (103) at (24, 3) {};
		\node [style=none] (106) at (26.25, 1) {};
		\node [style=none] (107) at (25.25, 1) {};
		\node [style=none] (109) at (22, 1) {};
		\node [style=none] (116) at (24.5, 1.5) {};
		\node [style=none] (118) at (22.75, 2) {};
		\node [style=none] (119) at (23.75, 2) {};
		\node [style=none] (126) at (37, 1.25) {};
		\node [style=none] (127) at (39.75, 0) {};
		\node [style=none] (128) at (42, 0) {};
		\node [style=none] (129) at (41, 0) {};
		\node [style=none] (130) at (41, 1) {};
		\node [style=none] (131) at (42, 1) {};
		\node [style=none] (133) at (42.75, 1) {};
		\node [style=none] (134) at (30.25, 1) {};
		\node [style=none] (135) at (39.5, 0) {};
		\node [style=none] (136) at (38.5, 0) {};
		\node [style=none] (137) at (38.5, 2) {};
		\node [style=none] (138) at (39.5, 2) {};
		\node [style=none] (139) at (40.75, 0) {};
		\node [style=none] (140) at (39.75, 0) {};
		\node [style=none] (141) at (39.75, 3) {};
		\node [style=none] (142) at (40.75, 3) {};
		\node [style=none] (143) at (39.25, 0) {};
		\node [style=none] (144) at (38.25, 0) {};
		\node [style=none] (145) at (37.25, 0) {};
		\node [style=none] (146) at (37.25, 1) {};
		\node [style=none] (147) at (38.25, 1) {};
		\node [style=none] (148) at (37, 0) {};
		\node [style=none] (149) at (36, 0) {};
		\node [style=none] (150) at (36, 1.75) {};
		\node [style=none] (151) at (37, 1.75) {};
		\node [style=none] (154) at (39.75, 3) {};
		\node [style=none] (155) at (33, 1) {};
		\node [style=none] (156) at (34, 1) {};
		\node [style=none] (157) at (34, 0) {};
		\node [style=none] (158) at (33, 0) {};
		\node [style=none] (160) at (37.75, 1) {};
		\node [style=none] (167) at (38.5, 2) {};
		\node [style=none] (168) at (39.5, 2) {};
		\node [style=none] (169) at (33, 1) {};
		\node [style=none] (170) at (34, 1) {};
		\node [style=none] (176) at (42.5, 0) {};
		\node [style=none] (177) at (30, 0) {};
		\node [style=none] (182) at (34.25, 1) {};
		\node [style=none] (183) at (35.25, 1) {};
		\node [style=none] (184) at (35.25, 0) {};
		\node [style=none] (185) at (34.25, 0) {};
		\node [style=none] (188) at (34.25, 1) {};
		\node [style=none] (189) at (35.25, 1) {};
		\node [style=none] (190) at (31.75, 1) {};
		\node [style=none] (191) at (32.75, 1) {};
		\node [style=none] (192) at (32.75, 0) {};
		\node [style=none] (193) at (31.75, 0) {};
		\node [style=none] (196) at (31.75, 1) {};
		\node [style=none] (197) at (30.5, 1) {};
		\node [style=none] (198) at (31.5, 1) {};
		\node [style=none] (199) at (31.5, 0) {};
		\node [style=none] (200) at (30.5, 0) {};
		\node [style=none] (203) at (30.5, 1) {};
		\node [style=none] (213) at (20.75, 1) {};
		\node [style=none] (218) at (40.25, 2) {};
		\node [style=none] (224) at (5.25, -1.25) {$H(\X)$};
		\node [style=none] (225) at (20.5, -1.25) {$H(\hat{\X})$};
		\node [style=none] (226) at (36.5, -1.25) {$H(\X')$};
		\node [style=none] (228) at (39.5, 2) {};
		\node [style=none] (229) at (38.5, 2) {};
		\node [style=none] (232) at (42, 1) {};
		\node [style=none] (233) at (41, 1) {};
		\node [style=none] (235) at (22.75, 2) {};
		\node [style=none] (236) at (23.75, 2) {};
		\node [style=none] (238) at (22.75, 2) {};
		\node [style=none] (239) at (23.75, 2) {};
		\node [style=none] (240) at (39.5, 3) {};
		\node [style=none] (241) at (39.5, 2) {};
		\node [style=none] (242) at (38.5, 2) {};
		\node [style=none] (243) at (38.5, 3) {};
		\node [style=none] (244) at (25.25, 1) {};
		\node [style=none] (245) at (26.25, 1) {};
		\node [style=none] (246) at (25.75, 1) {};
		\node [style=none] (247) at (25.25, 1) {};
		\node [style=none] (248) at (26.25, 1) {};
		\node [style=none] (249) at (42, 2) {};
		\node [style=none] (250) at (42, 1) {};
		\node [style=none] (251) at (41, 1) {};
		\node [style=none] (252) at (41, 2) {};
		\node [style=none] (253) at (13.75, 1) {};
		\node [style=none] (254) at (26.5, 1.5) {};
		\node [style=none] (255) at (26.5, -0.25) {};
		\node [style=none] (256) at (13.75, -0.25) {};
		\node [style=none] (258) at (3.5, 2.5) {};
		\node [style=none] (259) at (10.25, 2.5) {};
		\node [style=none] (260) at (10.25, -0.25) {};
		\node [style=none] (261) at (3.5, -0.25) {};
		\node [style=none] (262) at (5, 3) {$U$};
		\node [style=none] (263) at (32.75, 2.75) {$U'$};
		\node [style=none] (264) at (29.75, 1) {};
		\node [style=none] (265) at (42.25, 1.5) {};
		\node [style=none] (266) at (42.25, -0.25) {};
		\node [style=none] (267) at (29.75, -0.25) {};
		\node [style=none] (270) at (27.5, 1.75) {};
		\node [style=none] (271) at (29, 1.75) {};
		\node [style=none] (272) at (43.25, 1) {$1$};
		\node [style=none] (274) at (28.25, 2.5) {$\deltaone$};
		\node [style=none] (275) at (12.25, 2.5) {$\deltatwo$};
		\node [style=none] (276) at (11.5, 1.75) {};
		\node [style=none] (277) at (13, 1.75) {};
		\node [style=none] (278) at (17, 1) {};
		\node [style=none] (279) at (18, 1) {};
		\node [style=none] (280) at (18, 0) {};
		\node [style=none] (281) at (17, 0) {};
		\node [style=none] (282) at (17, 1) {};
		\node [style=none] (283) at (18, 1) {};
		\node [style=none] (284) at (18.25, 1) {};
		\node [style=none] (285) at (19.25, 1) {};
		\node [style=none] (286) at (19.25, 0) {};
		\node [style=none] (287) at (18.25, 0) {};
		\node [style=none] (288) at (18.25, 1) {};
		\node [style=none] (289) at (19.25, 1) {};
		\node [style=none] (290) at (15.75, 1) {};
		\node [style=none] (291) at (16.75, 1) {};
		\node [style=none] (292) at (16.75, 0) {};
		\node [style=none] (293) at (15.75, 0) {};
		\node [style=none] (294) at (15.75, 1) {};
		\node [style=none] (295) at (14.5, 1) {};
		\node [style=none] (296) at (15.5, 1) {};
		\node [style=none] (297) at (15.5, 0) {};
		\node [style=none] (298) at (14.5, 0) {};
		\node [style=none] (299) at (14.5, 1) {};
		\node [style=none] (300) at (16.5, 2.75) {$U'$};
	\end{pgfonlayer}
	\begin{pgfonlayer}{edgelayer}
		\draw [style=FILLGREEN] (3.center) to (4.center);
		\draw [style=FILLBLUE] (7.center)
			 to (6.center)
			 to (5.center)
			 to (8.center)
			 to cycle;
		\draw [style=thick] (0.center) to (9.center);
		\draw [style=FILLBLUE] (13.center)
			 to (12.center)
			 to (15.center)
			 to (14.center)
			 to cycle;
		\draw [style=FILLBLUE] (17.center)
			 to (16.center)
			 to (19.center)
			 to (18.center)
			 to cycle;
		\draw [style=FILLBLUE] (21.center)
			 to (20.center)
			 to (23.center)
			 to (22.center)
			 to cycle;
		\draw [style=FILLBLUE] (25.center)
			 to (24.center)
			 to (27.center)
			 to (26.center)
			 to cycle;
		\draw [style=FILLGREEN] (34.center) to (35.center);
		\draw [style=FILLGREEN] (38.center) to (39.center);
		\draw [style=FILLGREEN] (42.center) to (43.center);
		\draw [style=EDGEDASHED] (11.center) to (10.center);
		\draw [style=FILLBLUE] (54.center) to (55.center);
		\draw [style=FILLBLUE] (71.center)
			 to (70.center)
			 to (69.center)
			 to (72.center)
			 to cycle;
		\draw [style=thick] (64.center) to (73.center);
		\draw [style=FILLBLUE] (77.center)
			 to (76.center)
			 to (79.center)
			 to (78.center)
			 to cycle;
		\draw [style=FILLBLUE] (81.center)
			 to (80.center)
			 to (83.center)
			 to (82.center)
			 to cycle;
		\draw [style=FILLBLUE] (85.center)
			 to (84.center)
			 to (87.center)
			 to (86.center)
			 to cycle;
		\draw [style=FILLBLUE] (89.center)
			 to (88.center)
			 to (91.center)
			 to (90.center)
			 to cycle;
		\draw [style=FILLGREEN] (102.center) to (103.center);
		\draw [style=FILLGREEN] (106.center) to (107.center);
		\draw [style=EDGEDASHED] (75.center) to (74.center);
		\draw [style=FILLBLUE] (118.center) to (119.center);
		\draw [style=FILLBLUE] (130.center)
			 to (129.center)
			 to (128.center)
			 to (131.center)
			 to cycle;
		\draw [style=FILLBLUE] (136.center)
			 to (135.center)
			 to (138.center)
			 to (137.center)
			 to cycle;
		\draw [style=FILLBLUE] (140.center)
			 to (139.center)
			 to (142.center)
			 to (141.center)
			 to cycle;
		\draw [style=FILLBLUE] (145.center)
			 to (144.center)
			 to (147.center)
			 to (146.center)
			 to cycle;
		\draw [style=FILLBLUE] (149.center)
			 to (148.center)
			 to (151.center)
			 to (150.center)
			 to cycle;
		\draw [style=FILLGREEN] (157.center)
			 to (158.center)
			 to (155.center)
			 to (156.center)
			 to cycle;
		\draw [style=EDGEDASHED] (134.center) to (133.center);
		\draw [style=FILLBLUE] (167.center) to (168.center);
		\draw [style=FILLBLUE] (169.center) to (170.center);
		\draw [style=thick] (176.center) to (177.center);
		\draw [style=FILLGREEN] (183.center)
			 to (184.center)
			 to (185.center)
			 to (182.center);
		\draw [style=FILLGREEN] (182.center) to (183.center);
		\draw [style=FILLGREEN] (191.center)
			 to (192.center)
			 to (193.center)
			 to (190.center);
		\draw [style=FILLGREEN] (190.center) to (191.center);
		\draw [style=FILLGREEN] (198.center)
			 to (199.center)
			 to (200.center)
			 to (197.center);
		\draw [style=FILLGREEN] (197.center) to (198.center);
		\draw [style=FILLGREEN] (229.center) to (228.center);
		\draw [style=FILLGREEN] (233.center) to (232.center);
		\draw [style=FILLBLUE] (236.center) to (235.center);
		\draw [style=FILLBLUE] (238.center) to (239.center);
		\draw [style=FILLGREEN] (242.center)
			 to (241.center)
			 to (240.center)
			 to (243.center)
			 to cycle;
		\draw [style=FILLBLUE] (245.center) to (244.center);
		\draw [style=FILLBLUE] (247.center) to (248.center);
		\draw [style=FILLGREEN] (251.center)
			 to (250.center)
			 to (249.center)
			 to (252.center)
			 to cycle;
		\draw [style=black dotted] (254.center) to (255.center);
		\draw [style=black dotted, bend right=360, looseness=0.75] (255.center) to (256.center);
		\draw [style=black dotted] (256.center) to (253.center);
		\draw [style=black dotted, bend left=90, looseness=0.75] (253.center) to (254.center);
		\draw [style=black dotted] (258.center)
			 to [bend left=90, looseness=0.75] (259.center)
			 to (260.center)
			 to [bend right=360, looseness=0.75] (261.center)
			 to cycle;
		\draw [style=black dotted] (265.center) to (266.center);
		\draw [style=black dotted, bend right=360, looseness=0.75] (266.center) to (267.center);
		\draw [style=black dotted] (267.center) to (264.center);
		\draw [style=black dotted, bend left=90, looseness=0.75] (264.center) to (265.center);
		\draw [style=new edge style 1] (270.center) to (271.center);
		\draw [style=new edge style 1] (276.center) to (277.center);
		\draw [style=FILLGREEN] (280.center)
			 to (281.center)
			 to (278.center)
			 to (279.center)
			 to cycle;
		\draw [style=FILLBLUE] (282.center) to (283.center);
		\draw [style=FILLGREEN] (285.center)
			 to (286.center)
			 to (287.center)
			 to (284.center);
		\draw [style=FILLGREEN] (284.center) to (285.center);
		\draw [style=FILLGREEN] (291.center)
			 to (292.center)
			 to (293.center)
			 to (290.center);
		\draw [style=FILLGREEN] (290.center) to (291.center);
		\draw [style=FILLGREEN] (296.center)
			 to (297.center)
			 to (298.center)
			 to (295.center);
		\draw [style=FILLGREEN] (295.center) to (296.center);
	\end{pgfonlayer}
\end{tikzpicture}

%% file: fig/intermediate-dataset-remove.tikz
\begin{tikzpicture}
	\begin{pgfonlayer}{nodelayer}
		\node [style=none] (0) at (42, 0) {};
		\node [style=none] (3) at (36.25, 2) {};
		\node [style=none] (4) at (35.25, 2) {};
		\node [style=none] (5) at (41.25, 0) {};
		\node [style=none] (6) at (40.25, 0) {};
		\node [style=none] (7) at (40.25, 1) {};
		\node [style=none] (8) at (41.25, 1) {};
		\node [style=none] (9) at (29.5, 0) {};
		\node [style=none] (10) at (42, 1) {};
		\node [style=none] (11) at (29.75, 1) {};
		\node [style=none] (12) at (38.75, 0) {};
		\node [style=none] (13) at (37.75, 0) {};
		\node [style=none] (14) at (37.75, 2) {};
		\node [style=none] (15) at (38.75, 2) {};
		\node [style=none] (16) at (40, 0) {};
		\node [style=none] (17) at (39, 0) {};
		\node [style=none] (18) at (39, 3) {};
		\node [style=none] (19) at (40, 3) {};
		\node [style=none] (20) at (37.5, 0) {};
		\node [style=none] (21) at (36.5, 0) {};
		\node [style=none] (22) at (36.5, 1) {};
		\node [style=none] (23) at (37.5, 1) {};
		\node [style=none] (24) at (36.25, 0) {};
		\node [style=none] (25) at (35.25, 0) {};
		\node [style=none] (26) at (35.25, 2) {};
		\node [style=none] (27) at (36.25, 2) {};
		\node [style=none] (31) at (42.5, 1) {$1$};
		\node [style=none] (34) at (37.5, 1) {};
		\node [style=none] (35) at (36.5, 1) {};
		\node [style=none] (38) at (40, 3) {};
		\node [style=none] (39) at (39, 3) {};
		\node [style=none] (42) at (41.25, 1) {};
		\node [style=none] (43) at (40.25, 1) {};
		\node [style=none] (48) at (38.25, 2) {};
		\node [style=none] (54) at (37.75, 2) {};
		\node [style=none] (55) at (38.75, 2) {};
		\node [style=none] (64) at (26.75, 0) {};
		\node [style=none] (69) at (26.25, 0) {};
		\node [style=none] (70) at (25.25, 0) {};
		\node [style=none] (71) at (25.25, 1) {};
		\node [style=none] (72) at (26.25, 1) {};
		\node [style=none] (73) at (13.75, 0) {};
		\node [style=none] (74) at (27, 1) {};
		\node [style=none] (75) at (14, 1) {};
		\node [style=none] (76) at (23.75, 0) {};
		\node [style=none] (77) at (22.75, 0) {};
		\node [style=none] (78) at (22.75, 2) {};
		\node [style=none] (79) at (23.75, 2) {};
		\node [style=none] (80) at (25, 0) {};
		\node [style=none] (81) at (24, 0) {};
		\node [style=none] (82) at (24, 3) {};
		\node [style=none] (83) at (25, 3) {};
		\node [style=none] (84) at (22.5, 0) {};
		\node [style=none] (85) at (21.5, 0) {};
		\node [style=none] (86) at (21.5, 1) {};
		\node [style=none] (87) at (22.5, 1) {};
		\node [style=none] (88) at (21.25, 0) {};
		\node [style=none] (89) at (20.25, 0) {};
		\node [style=none] (90) at (20.25, 1.75) {};
		\node [style=none] (91) at (21.25, 1.75) {};
		\node [style=none] (95) at (27.5, 1) {$1$};
		\node [style=none] (102) at (25, 3) {};
		\node [style=none] (103) at (24, 3) {};
		\node [style=none] (106) at (26.25, 1) {};
		\node [style=none] (107) at (25.25, 1) {};
		\node [style=none] (109) at (22, 1) {};
		\node [style=none] (116) at (24.5, 1.5) {};
		\node [style=none] (118) at (22.75, 2) {};
		\node [style=none] (119) at (23.75, 2) {};
		\node [style=none] (126) at (5.5, 1.25) {};
		\node [style=none] (127) at (8.25, 0) {};
		\node [style=none] (128) at (10.5, 0) {};
		\node [style=none] (129) at (9.5, 0) {};
		\node [style=none] (130) at (9.5, 1) {};
		\node [style=none] (131) at (10.5, 1) {};
		\node [style=none] (133) at (11.25, 1) {};
		\node [style=none] (134) at (-1.25, 1) {};
		\node [style=none] (135) at (8, 0) {};
		\node [style=none] (136) at (7, 0) {};
		\node [style=none] (137) at (7, 2) {};
		\node [style=none] (138) at (8, 2) {};
		\node [style=none] (139) at (9.25, 0) {};
		\node [style=none] (140) at (8.25, 0) {};
		\node [style=none] (141) at (8.25, 3) {};
		\node [style=none] (142) at (9.25, 3) {};
		\node [style=none] (143) at (7.75, 0) {};
		\node [style=none] (144) at (6.75, 0) {};
		\node [style=none] (145) at (5.75, 0) {};
		\node [style=none] (146) at (5.75, 1) {};
		\node [style=none] (147) at (6.75, 1) {};
		\node [style=none] (148) at (5.5, 0) {};
		\node [style=none] (149) at (4.5, 0) {};
		\node [style=none] (150) at (4.5, 1.75) {};
		\node [style=none] (151) at (5.5, 1.75) {};
		\node [style=none] (154) at (8.25, 3) {};
		\node [style=none] (155) at (1.5, 1) {};
		\node [style=none] (156) at (2.5, 1) {};
		\node [style=none] (157) at (2.5, 0) {};
		\node [style=none] (158) at (1.5, 0) {};
		\node [style=none] (160) at (6.25, 1) {};
		\node [style=none] (167) at (7, 2) {};
		\node [style=none] (168) at (8, 2) {};
		\node [style=none] (169) at (1.5, 1) {};
		\node [style=none] (170) at (2.5, 1) {};
		\node [style=none] (176) at (11, 0) {};
		\node [style=none] (177) at (-1.5, 0) {};
		\node [style=none] (182) at (2.75, 1) {};
		\node [style=none] (183) at (3.75, 1) {};
		\node [style=none] (184) at (3.75, 0) {};
		\node [style=none] (185) at (2.75, 0) {};
		\node [style=none] (188) at (2.75, 1) {};
		\node [style=none] (189) at (3.75, 1) {};
		\node [style=none] (190) at (0.25, 1) {};
		\node [style=none] (191) at (1.25, 1) {};
		\node [style=none] (192) at (1.25, 0) {};
		\node [style=none] (193) at (0.25, 0) {};
		\node [style=none] (196) at (0.25, 1) {};
		\node [style=none] (197) at (-1, 1) {};
		\node [style=none] (198) at (0, 1) {};
		\node [style=none] (199) at (0, 0) {};
		\node [style=none] (200) at (-1, 0) {};
		\node [style=none] (203) at (-1, 1) {};
		\node [style=none] (213) at (20.75, 1) {};
		\node [style=none] (218) at (8.75, 2) {};
		\node [style=none] (224) at (36.5, -1.25) {$H(\X')$};
		\node [style=none] (225) at (20.5, -1.25) {$H(\hat{\X})$};
		\node [style=none] (226) at (4.75, -1.25) {$H(\X)$};
		\node [style=none] (228) at (8, 2) {};
		\node [style=none] (229) at (7, 2) {};
		\node [style=none] (232) at (10.5, 1) {};
		\node [style=none] (233) at (9.5, 1) {};
		\node [style=none] (235) at (22.75, 2) {};
		\node [style=none] (236) at (23.75, 2) {};
		\node [style=none] (238) at (22.75, 2) {};
		\node [style=none] (239) at (23.75, 2) {};
		\node [style=none] (240) at (8, 3) {};
		\node [style=none] (241) at (8, 2) {};
		\node [style=none] (242) at (7, 2) {};
		\node [style=none] (243) at (7, 3) {};
		\node [style=none] (244) at (25.25, 1) {};
		\node [style=none] (245) at (26.25, 1) {};
		\node [style=none] (246) at (25.75, 1) {};
		\node [style=none] (247) at (25.25, 1) {};
		\node [style=none] (248) at (26.25, 1) {};
		\node [style=none] (249) at (10.5, 2) {};
		\node [style=none] (250) at (10.5, 1) {};
		\node [style=none] (251) at (9.5, 1) {};
		\node [style=none] (252) at (9.5, 2) {};
		\node [style=none] (258) at (34.75, 2.5) {};
		\node [style=none] (259) at (41.5, 2.5) {};
		\node [style=none] (260) at (41.5, -0.25) {};
		\node [style=none] (261) at (34.75, -0.25) {};
		\node [style=none] (262) at (36.25, 3) {$U'$};
		\node [style=none] (263) at (1.25, 2.75) {$U$};
		\node [style=none] (264) at (-1.75, 1) {};
		\node [style=none] (265) at (10.75, 1.5) {};
		\node [style=none] (266) at (10.75, -0.25) {};
		\node [style=none] (267) at (-1.75, -0.25) {};
		\node [style=none] (270) at (27.5, 1.75) {};
		\node [style=none] (271) at (29, 1.75) {};
		\node [style=none] (272) at (11.75, 1) {$1$};
		\node [style=none] (274) at (28.25, 2.5) {$\deltaone$};
		\node [style=none] (275) at (12.25, 2.5) {$\deltatwo$};
		\node [style=none] (276) at (11.5, 1.75) {};
		\node [style=none] (277) at (13, 1.75) {};
		\node [style=none] (300) at (20.75, 2.75) {$U'$};
		\node [style=none] (301) at (23.75, 3) {};
		\node [style=none] (302) at (23.75, 2) {};
		\node [style=none] (303) at (22.75, 2) {};
		\node [style=none] (304) at (22.75, 3) {};
		\node [style=none] (305) at (26.25, 2) {};
		\node [style=none] (306) at (26.25, 1) {};
		\node [style=none] (307) at (25.25, 1) {};
		\node [style=none] (308) at (25.25, 2) {};
		\node [style=none] (309) at (19.75, 2.5) {};
		\node [style=none] (310) at (26.5, 2.5) {};
		\node [style=none] (311) at (26.5, -0.25) {};
		\node [style=none] (312) at (19.75, -0.25) {};
	\end{pgfonlayer}
	\begin{pgfonlayer}{edgelayer}
		\draw [style=FILLGREEN] (3.center) to (4.center);
		\draw [style=FILLBLUE] (7.center)
			 to (6.center)
			 to (5.center)
			 to (8.center)
			 to cycle;
		\draw [style=thick] (0.center) to (9.center);
		\draw [style=FILLBLUE] (13.center)
			 to (12.center)
			 to (15.center)
			 to (14.center)
			 to cycle;
		\draw [style=FILLBLUE] (17.center)
			 to (16.center)
			 to (19.center)
			 to (18.center)
			 to cycle;
		\draw [style=FILLBLUE] (21.center)
			 to (20.center)
			 to (23.center)
			 to (22.center)
			 to cycle;
		\draw [style=FILLBLUE] (25.center)
			 to (24.center)
			 to (27.center)
			 to (26.center)
			 to cycle;
		\draw [style=FILLGREEN] (34.center) to (35.center);
		\draw [style=FILLGREEN] (38.center) to (39.center);
		\draw [style=FILLGREEN] (42.center) to (43.center);
		\draw [style=EDGEDASHED] (11.center) to (10.center);
		\draw [style=FILLBLUE] (54.center) to (55.center);
		\draw [style=FILLBLUE] (71.center)
			 to (70.center)
			 to (69.center)
			 to (72.center)
			 to cycle;
		\draw [style=thick] (64.center) to (73.center);
		\draw [style=FILLBLUE] (77.center)
			 to (76.center)
			 to (79.center)
			 to (78.center)
			 to cycle;
		\draw [style=FILLBLUE] (81.center)
			 to (80.center)
			 to (83.center)
			 to (82.center)
			 to cycle;
		\draw [style=FILLBLUE] (85.center)
			 to (84.center)
			 to (87.center)
			 to (86.center)
			 to cycle;
		\draw [style=FILLBLUE] (89.center)
			 to (88.center)
			 to (91.center)
			 to (90.center)
			 to cycle;
		\draw [style=FILLGREEN] (102.center) to (103.center);
		\draw [style=FILLGREEN] (106.center) to (107.center);
		\draw [style=EDGEDASHED] (75.center) to (74.center);
		\draw [style=FILLBLUE] (118.center) to (119.center);
		\draw [style=FILLBLUE] (130.center)
			 to (129.center)
			 to (128.center)
			 to (131.center)
			 to cycle;
		\draw [style=FILLBLUE] (136.center)
			 to (135.center)
			 to (138.center)
			 to (137.center)
			 to cycle;
		\draw [style=FILLBLUE] (140.center)
			 to (139.center)
			 to (142.center)
			 to (141.center)
			 to cycle;
		\draw [style=FILLBLUE] (145.center)
			 to (144.center)
			 to (147.center)
			 to (146.center)
			 to cycle;
		\draw [style=FILLBLUE] (149.center)
			 to (148.center)
			 to (151.center)
			 to (150.center)
			 to cycle;
		\draw [style=FILLRED] (157.center)
			 to (158.center)
			 to (155.center)
			 to (156.center)
			 to cycle;
		\draw [style=EDGEDASHED] (134.center) to (133.center);
		\draw [style=FILLBLUE] (167.center) to (168.center);
		\draw [style=FILLRED] (169.center) to (170.center);
		\draw [style=thick] (176.center) to (177.center);
		\draw [style=FILLRED] (183.center)
			 to (184.center)
			 to (185.center)
			 to (182.center);
		\draw [style=FILLGREEN] (182.center) to (183.center);
		\draw [style=FILLRED] (191.center)
			 to (192.center)
			 to (193.center)
			 to (190.center);
		\draw [style=FILLGREEN] (190.center) to (191.center);
		\draw [style=FILLRED] (198.center)
			 to (199.center)
			 to (200.center)
			 to (197.center);
		\draw [style=FILLGREEN] (197.center) to (198.center);
		\draw [style=FILLGREEN] (229.center) to (228.center);
		\draw [style=FILLGREEN] (233.center) to (232.center);
		\draw [style=FILLBLUE] (236.center) to (235.center);
		\draw [style=FILLBLUE] (238.center) to (239.center);
		\draw [style=FILLRED] (242.center)
			 to (241.center)
			 to (240.center)
			 to (243.center)
			 to cycle;
		\draw [style=FILLBLUE] (245.center) to (244.center);
		\draw [style=FILLBLUE] (247.center) to (248.center);
		\draw [style=FILLRED] (251.center)
			 to (250.center)
			 to (249.center)
			 to (252.center)
			 to cycle;
		\draw [style=black dotted] (258.center)
			 to [bend left=90, looseness=0.75] (259.center)
			 to (260.center)
			 to [bend right=360, looseness=0.75] (261.center)
			 to cycle;
		\draw [style=black dotted] (265.center) to (266.center);
		\draw [style=black dotted, bend right=360, looseness=0.75] (266.center) to (267.center);
		\draw [style=black dotted] (267.center) to (264.center);
		\draw [style=black dotted, bend left=90, looseness=0.75] (264.center) to (265.center);
		\draw [style=new edge style 1] (270.center) to (271.center);
		\draw [style=new edge style 1] (276.center) to (277.center);
		\draw [style=FILLRED] (303.center)
			 to (302.center)
			 to (301.center)
			 to (304.center)
			 to cycle;
		\draw [style=FILLRED] (307.center)
			 to (306.center)
			 to (305.center)
			 to (308.center)
			 to cycle;
		\draw [style=black dotted] (309.center)
			 to [bend left=90, looseness=0.75] (310.center)
			 to (311.center)
			 to [bend right=360, looseness=0.75] (312.center)
			 to cycle;
	\end{pgfonlayer}
\end{tikzpicture}

%% file: content/top-k.tex
\section{Top-k Counting Queries}
\label{sec:top-k}
The privacy guarantees of \Cref{alg:main} are conditioned on the histogram being $k$-sparse for all datasets.
Here we present a technique when we do not have this guarantee.
Specifically, we consider the setting when the input is a dataset $\X = ({X}_1, \cdots, {X}_n)$ where ${X}_i \in \{0,1\}^d$.
We want to release a private estimate of $H(\X) = \sum_{i = 1}^n X_i$, and ${X}_i$ can have any number of non-zero entries. 
We use superscript to denote the elements of the histogram in descending order with ties broken arbitrarily such that $H(\X)^{(1)} \geq H(\X)^{(2)} \geq \dots \geq H(\X)^{(d - 1)} \geq H(\X)^{(d)}$.
This setting is studied in a line of work for \emph{Private top-k selection}:
\cite{durfee2019,ROSEN1997135,QiaoSZ21,mcsherry2017,bafna2017price,hao2024}.

Our algorithm in this setting relies on a simple pre-processing step.
We first find the value of the $(k + 1)$'th largest entry in the histogram.
We then subtract that value from all entries in the histogram and remove negative counts. 
This gives us a new histogram which we use as input for \Cref{alg:main}.
We show that this histogram is both $k$-sparse and monotonic which implies that the mechanism has the same privacy guarantees as \Cref{alg:main}.

\begin{algorithm}[H]
    \caption{Top-k Mechanism using Correlated Gaussian Noise}
    \label{alg:topk}
    \begin{algorithmic}[1]
        \Require Parameters $k$, $\sigma$, and $\tau$, dataset $\X = (X_1, \cdots, X_n)$ where $X_i \in \{0,1\}^d$. 
        \State Let $H(\X) = \sum_{i = 1}^n X_i$. %
        \State Let $\tilde{H}(\X) = \{0\}^d$.
        \For{each $i \in [d]$}
        \State Set $\tilde{H}(\X)_i = \max(0, H(\X)_i - H(\X)^{(k + 1)})$. \label{line:subtract-(k+1)}
        \EndFor
        \State \textbf{Release} $\algoname (k, \sigma, \tau, \tilde{H}(\X))$. %
    \end{algorithmic}
\end{algorithm}

\begin{restatable}{lemma}{monotonicity}
    \label{lem:top-k-sens-structure}
    The function $\tilde{H} : \mathcal{U}^{\mathbb{N}} \rightarrow \mathbb{N}^d$ in \Cref{alg:topk} produces $k$-sparse monotonic histograms.
\end{restatable}
 
\begin{proof}
    By \Cref{def:k-sparse-histogram} we have to show two properties of $\tilde{H}$. 
    It must hold for any $\X$ that $\|\tilde{H}(\X)\|_0 \leq k$ and for any pair $\X \sim \X'$ we have $\tilde{H}(\X) - \tilde{H}(\X') \in \{0,1\}^d$ or $\tilde{H}(\X) - \tilde{H}(\X') \in \{0,-1\}^d$.
    
    The sparsity claim is easy to see. 
    Any counter that is not strictly larger that $H(\X)^{k + 1}$ is removed in line \ref{line:subtract-(k+1)}. 
    By definition of $H(\X')^{k+1}$, there are at most $k$ such entries.

    To prove the monotonicity property, we must show that counters in $\Tilde{H}$ either all increase or all decrease by at most one.
    We only give the proof for the case where $\X$ is constructed by adding one data point to $\X'$. 
    The proof is symmetric for the case where $\X$ is created by removing one data point from $\X'$.

    We first partition $H(\X')$ into three sets: 
    Let $U = \{ i \in [d] \mid H(\X')_i > H(\X')^{(k + 1)} \}$,
    $M = \{i \in [d] \mid H(\X')_i = H(\X')^{(k + 1)}\}$ and $L = [d]\setminus \{U, M\}$. 
    Because we have $H(\X)_i - H(\X')_i \in \{0,1\}$ for all $i \in [d]$,  also $H(\X)^{(k + 1)} - H(\X')^{(k + 1)} \in \{0,1\}$.
    Note that the $(k+1)$'th largest entry might not represent the same element, but we only care about the value.
    
    Consider the case where $H(\X)^{(k + 1)} = H(\X')^{(k + 1)}$.
    We see that for all $l \in L: \tilde{H}(\X)_l = \tilde{H}(\X')_l = 0$, and for all $u\in U$ we have $\tilde{H}(\X')_u - \tilde{H}(\X)_u = H(\X')_u -H(\X)^{(k + 1)} - H(\X)_u + H(\X')^{(k + 1)} \in \{0, 1\}$ because we subtract the same value and increment $H(\X')_u$ by at most one. %
    Some elements $m \in M$ might now be exactly one larger than $H(\X')^{(k + 1)}$ in $H(\X)$, but still $\tilde{H}(\X)_m \in \{0, 1\}$ by definition.
    Thus we have that $\tilde{H}(\X) - \tilde{H}(\X')  \in \{0, 1\}^d$.
    
    Now, consider the case where $H(\X)^{(k + 1)} = H(\X')^{(k + 1)} + 1$.
    No element $l \in L$ can become larger or equal to the new $(k+1)$'th largest element, so still $\tilde{H}(\X')_l = \tilde{H}(\X)_l = 0$.
    Elements in $u \in U$ get reduced by at most one for $H(\X)$, because they either are increased together with the $H(\X')^{(k + 1)}$ and then $\tilde{H}(\X)_u - \tilde{H}(\X')_u = 0$, or they stay the same in which case $\tilde{H}(\X)_u = \tilde{H}(\X')_u - 1$.
    One can also see that for $m \in M$, ${H}(\X)_m - H(\X) ^{(k + 1)}\in \{0, -1\}$, depending on whether or not they increase together with $H(\X)^{(k + 1)}$. As such we have $\tilde{H}(\X)_m = 0$ for all $m \in M$.
    Therefore, $\tilde{H}(\X) - \tilde{H}(\X')  \in \{0, -1\}^d$.
\end{proof}

Since \Cref{alg:topk} simply returns the output of running \Cref{alg:main} with a $k$-sparse monotonic histogram it has the same privacy guarantees.

\begin{corollary}
    \label{cor:top-k-privacy}
    The privacy guarantees specified by \Cref{thm:algorithm-privacy-add-the-deltas} and \Cref{th:max-the-delta} hold for \Cref{alg:topk}.
\end{corollary}

\Cref{alg:topk} introduces bias when subtracting $H(\X)^{(k + 1)}$ from each counter as a pre-processing step. 
If we have access to a private estimate of $H(\X)^{(k + 1)}$ we can add it to all counters of the output as post-processing.
Since this estimate would be used for multiple counters, similar to $Z_{corr}$, we might want to use additional privacy budget to get a more accurate estimate.
This can be included directly in the privacy analysis.
If we e.g. release $\tilde{H}(\X)^{(k + 1)} = H(\X)^{(k + 1)} + \Nd(0, \sigma^2/\sqrt{k})$ the privacy guarantees from \Cref{thm:algorithm-privacy-add-the-deltas} holds if we change $\sqrt{k + \sqrt{k}}$ to $\sqrt{k + 5\sqrt{k}}$ in $\deltaone$.

%% file: content/empirical-eval.tex
\section{Experiments}\label{sec:empirical-eval}

To back up our theoretical claims, we compare the error of the \emph{Correlated Stability Histogram} against both the \atd~approach \cite{googlelibthreshold} and the \emph{exact} analysis \cite{Wilkins24-GaussianSparseHistogramMechanism} of the uncorrelated GSHM. 
We consider the error in both privacy analysis approaches for our mechanisms as well (\Cref{thm:algorithm-privacy-add-the-deltas,thm:privacy-tighter-analysis}).

\begin{figure}[ht]
  \centering
  \subfloat[\centering Same $\epsilon, \delta, k$ as in \citet{Wilkins24-GaussianSparseHistogramMechanism}]{{\includegraphics[width=0.46\linewidth]{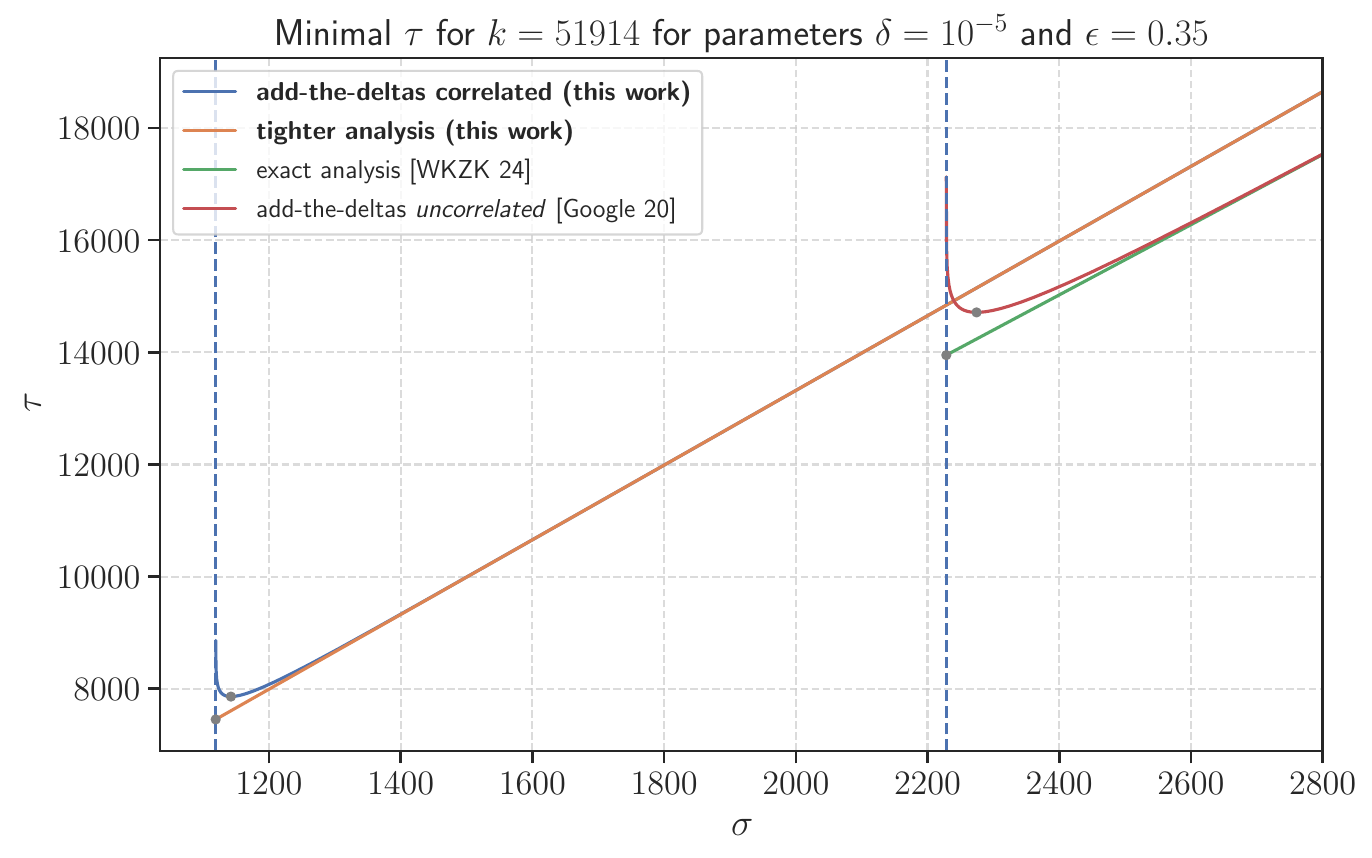} }}%
    \qquad
    \subfloat[\centering Some improvement also for smaller $k$.]{{\includegraphics[width=0.46\linewidth]{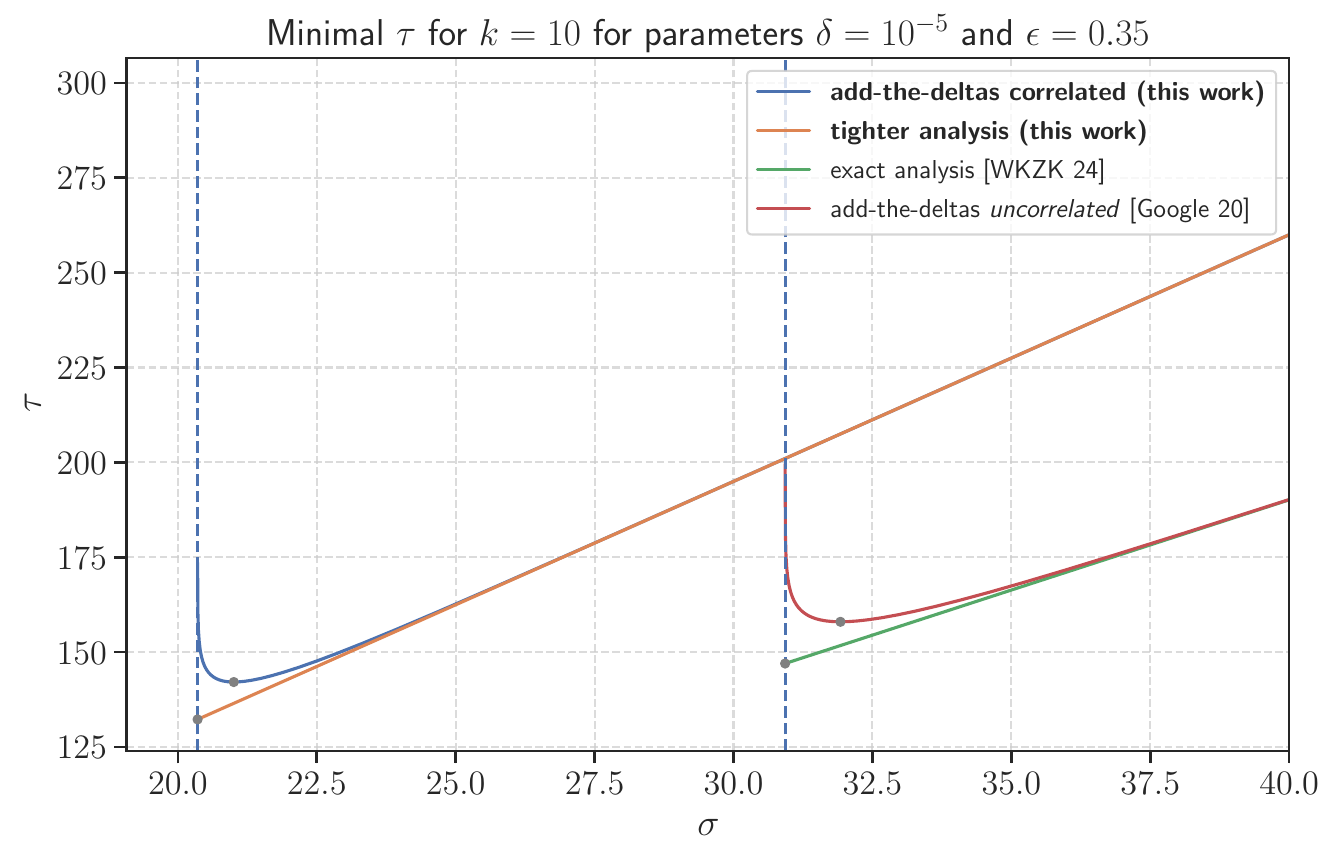} }}%
  \caption{The results of our experiments. Using the same parameters as in \cite{Wilkins24-GaussianSparseHistogramMechanism}, the graphs show the minimum $\tau$ required to get $(0.35, 10^{-5})$-DP guarantees for a noise level $\sigma$.
  The {\setulcolor{sbGreen}\ul{green}} line denotes the tight analysis of \citet{Wilkins24-GaussianSparseHistogramMechanism}, the {\setulcolor{sbRed}\ul{red}} shows the \atd~ \cite{googlelibthreshold} approach and the {\setulcolor{sbBlue}\ul{blue}} and {\setulcolor{sbOrange}\ul{orange}} lines are our results. %
  The marked points denote the minimum $\tau$ for each technique. %
  \textbf{a)} Uses the same parameters as in Wilkins. 
  As high values of $k$ are preferable for our mechanism, we bring down the threshold from $\approx13950$ to $\approx 7860$, lowering it by $\approx43\%$:
  \textbf{b)} We get some small improvement even for small $k$ values.
  Note that since our mechanism adds two noise samples the plot shows the total magnitude of noise. %
  } \label{fig:experiments}%
\end{figure}

We compare the parameters of the mechanisms when releasing $k$-sparse monotonic histograms.
We ran the experiments shown in \cref{fig:experiments} with the same privacy parameters as in \citet{Wilkins24-GaussianSparseHistogramMechanism}\footnote{in turn based on the privacy guarantees of the \emph{Facebook URL dataset} \cite{DVN/TDOAPG_2020}.}, which is $\epsilon = 0.35$ and  $\delta = 10^{-5}$.
Following their approach, we plot the minimum $\tau$ such that each mechanism satisfies $(\varepsilon,\delta)$-DP for a given magnitude of noise.
Note that our setting of $\|H(\X)\|_0 \leq k$ differs from \cite{Wilkins24-GaussianSparseHistogramMechanism}, so the experiments are not directly comparable. %

\paragraph{Results.} The \emph{first plot} shown in \cref{fig:experiments}\textbf{a)} resembles the same $k = 51914$ used in \cite[Figure~1~(A)]{Wilkins24-GaussianSparseHistogramMechanism}.
In this setting, one can lower the threshold by approximately $43\%$.
Because our technique favors large $k$ (we have to scale the noise with $\frac{1}{2}\sqrt{k + \sqrt{k}}$ instead of $\sqrt{k}$), we also include the case were $k=10$ is small in \cref{fig:experiments}\textbf{b)}.
The plots show that we make some small improvements even in that case.
Even in that case, our looser \atd~ analysis for the \algoname~ beats~\citet{Wilkins24-GaussianSparseHistogramMechanism}.
Note that the values of $\sigma$ in \Cref{alg:uncorrelated-GSHM} and \Cref{alg:main} are not directly comparable, because we add noise twice in \Cref{alg:main}.
For a fair comparison, we instead use the total magnitude of noise in the plots. 
For the CSH we plot the value of $(\sigma^2 + \sigma^2/\sqrt{k})^{1/2} = (1 + 1/\sqrt{k})^{1/2}\sigma$.
The dotted lines indicated the minimum magnitude of total noise for which GSHM and CSH satisfy $(\varepsilon, \delta)$-DP.

%% file: content/aggregators.tex
\section{Extensions}
\label{sec:aggregators}

The setting in \citet{Wilkins24-GaussianSparseHistogramMechanism} is slightly different from ours.
We will discuss how to adapt our technique to their setting and consider two extensions of the GSHM we did not include in the pseudocode of \Cref{alg:uncorrelated-GSHM}.

\paragraph{Additional sparsity threshold.}
The mechanism of \citet{Wilkins24-GaussianSparseHistogramMechanism} employs a second threshold $\tau' < \tau$ that allows to filter out infrequent data in a pre-processing step: All counters below $\tau'$ are removed before adding noise.
The high threshold $\tau$ is then used to remove noisy counters similar to both \Cref{alg:uncorrelated-GSHM,alg:main}.
This generalized setup allows them to account for pre-processing steps that the privacy expert has no control over.
Our setting corresponds to the case where $\tau' = 1$, and it is straightforward to incorporate this constraint into our mechanism:
For a given $\tau'$, we simply have to replace $\tau$ with the difference between the two thresholds in all theorems.
In fact, this situation is very similar to our result in \Cref{sec:top-k} where we remove all values below a lower threshold $\tau' = 1 + H(\vec{X})^{(k + 1)}$ before adding noise.

Note that the lower threshold of \citet{Wilkins24-GaussianSparseHistogramMechanism} is assumed to be data independent.
If it is, one must take care not to violate privacy.
The privacy guarantees hold if we can give guarantees about the structure of pre-processed neighboring datasets similar to \Cref{lem:top-k-sens-structure}.

\paragraph{Aggregator functions.}
\citet{Wilkins24-GaussianSparseHistogramMechanism} consider a setting where we are given an aggregator function $\cA$ which returns a vector when applied to any dataset $\X$.

If any $j \in [d]$ in the privatized released histogram $\tilde{H}(\X)_j$, exceeds the threshold, the aggregator function is applied to all data points where $\X_{i,j} = 1$.
The aggregated values are then privatized by adding Gaussian noise which shape and magnitude can be set independently of the magnitude of noise added to the count.
The privatized aggregate is then released along with $\tilde{H}(\X)_j$.
This setup is motivated by \emph{group-by} database operations where we first want to privately estimate the count of each group together with some aggregates modeled by $\cA$.

We did not consider aggregate functions in our analysis and instead focused on the classical setting, where we only wanted to privately release a sparse histogram while ignoring zero counters.
\citeauthor*{Wilkins24-GaussianSparseHistogramMechanism} accounts for the impact of the aggregator functions in the equivalent of $\deltaone$ in \cite[Theorem 5.4, Corollary 5.4.1]{Wilkins24-GaussianSparseHistogramMechanism}

In short, the effect of aggregators can be accounted for as an increase in the $\ell_2$ sensitivity using a transformation of the aggregate function.
The same technique could be applied to our mechanism. 
This would give us a similar improvement as \cref{lem:computing-threshold-add-the-delta} over the standard GSHM.
It's not as clear how the mechanisms compare under the tighter analyses.

The core idea behind the correlated noise can also be used to improve some aggregate queries. 
This can be used even in settings where we do not want to add correlated noise to the counts.
The mechanism of \citeauthor{lebeda2024} uses an estimate of the dataset size $\tilde{n}$ to reduce the independent noise added to each query. 
They spent part of the privacy budget to estimate $\tilde{n}$.
However, in our setting, we already have a private estimate for the dataset size for the aggregate query in $\tilde{H}(\X)_j$.

If each query is a \emph{sum} query where users have a value between $0$ and some value $C \in \mathbb{R}$, we can reduce the magnitude of noise by a factor of $2$ by adding $(\tilde{n} - n)C/2$ to each sum (It follows from \cite[Lemma~3.6]{lebeda2024}). 
We can use a similar trick if $\cA$ computes a sum of values in some range $[L, U]$. 
The standard approach would be to add noise with magnitude scaled to $\max(\vert L \vert, \vert U \vert)$
(see e.g. \cite[Table~1]{WilsonZLDSG20-SQL-bounded-contribution}).
Instead, we can recenter values around zero. 
As an example, $\cA$ contains a sum query with values in $[100,200]$. 
We subtract $(L + U)/2$ from each point so that we instead sum values in $[-50,50]$ and reduce the sensitivity from $200$ to $50$.
As post-processing, we add $150 \cdot \tilde{H}(\X)_j$ to the estimate of the new sum.
The error then depends on $\tilde{H}(\X)_j - H(\X)_j$.
As with the Correlated Gaussian Mechanism, we gain an advantage over estimating the sum directly if $\mathcal{A}$ contains multiple such sums because we can reuse the estimate $\tilde{H}(\X)_j$ for each sum.

%% file: content/discrete-gaussian.tex
\section{From Theory towards Practice: Discrete Gaussian Noise}\label{sec:discrete}
All mechanisms discussed in this paper achieve privacy guarantees using continuous Gaussian noise which is a standard tool in the differential privacy literature. 
However, real numbers cannot be represented exactly on computers which makes the implementation challenging.
We can instead use the Discrete Gaussian Mechanism due to Canonne, Kamath, and Steinke [\citeyear{Canonne_Kamath_Steinke_2022}].
In this chapter, we modify the technique by \citet{lebeda2024} to make the mechanism compatible with discrete Gaussian noise $ \discGauss(0, \sigma^2)$ and then provide privacy guarantees for the discrete analogue to \Cref{alg:main} using $\rho$-zCDP.
We first list required definitions and the privacy guarantees of the \emph{Multivariate Discrete Gaussian}:

\begin{definition}[Discrete Gaussian Distribution]
\label{def:discrete-distribution-gaussian}
Let $\sigma \in \mathbb{R}$ with $\sigma > 0$. The discrete Gaussian distribution with mean $0$ and scale $\sigma$ is denoted $\discGauss(0, \sigma^2)$. 
It is a probability distribution supported on the integers and defined by $\forall x \in \mathbb{Z}: \Pr\limits_{X \sim \discGauss(0, \sigma^2)}\left[X = x\right] = \dfrac{e^{-x^2/(2\sigma^2)}}{\sum_{y \in \mathbb{Z}} e^{-y^2/(2\sigma^2)}}$
\end{definition}

\begin{definition}[\cite{zeroConcentrated} Zero-Concentrated Differential Privacy]
    \label{def:zero-concentrated-differential-privacy}
Given $\rho > 0$, a randomized mechanism $\mathcal{M}: \mathcal{U}^{\mathbb{N}} \rightarrow \mathcal{Y}$ satisfies $\rho$-zCDP, if for every pair of neighboring datasets $\X \sim \X'$ and all $\alpha > 1$ it holds that
\[
    D_\alpha\left(\mathcal{M}(\X) \vert \vert \mathcal{M}(\X')\right) \leq \rho \alpha \,,
\]
where $D_\alpha(\mathcal{M}(\X) \vert \vert \mathcal{M}(\X'))$ denotes the $\alpha$-Rényi divergence between the two distributions $\mathcal{M}(\X)$ and $\mathcal{M}(\X')$.
Furthermore, Zero-Concentrated Differential Privacy is immune to post-processing.
\end{definition}

\begin{lemma}[\cite{zeroConcentrated} zCDP implies approximate DP]
    \label{lem:zCDP-implies-approx-DP}
    If a randomized mechanism $\mathcal{M}$ satisfies $\rho$-zCDP, then $\mathcal{M}$ is $(\varepsilon, \delta)$-DP for any $\delta > 0$ and $\varepsilon=\rho + 2\sqrt{\rho \log(1/\delta)}$.
\end{lemma}

\begin{lemma}[{\cite[Theorem 2.13]{Canonne_Kamath_Steinke_2022}} Multivariate Discrete Gaussian is zCDP ]
\label{lem:discrete-gaussian-mechanism}
Let $\sigma_1,\dots,\sigma_d > 0$ and $\rho > 0$. 
Let $q:\mathcal{U}^{\mathbb{N}} \rightarrow \mathbb{Z}^d$ satisfy $\sum_{i \in [d]} (q(\X)_i - q(\X')_i)^2/\sigma_i^2 \leq 2\rho$ for all neighboring datasets $\X \sim \X'$. 
Define a randomized algorithm $M: \mathcal{U}^{\mathbb{N}} \rightarrow \mathbb{Z}^d$ by $M(\X) = q(\X) + Z$ where $Z_i \sim \discGauss(0, \sigma_i^2)$ independently for each $i \in [d]$. 
Then $M$ satisfies $\rho$-zCDP.  
\end{lemma}


Next, we present a discrete variant of the Correlated Gaussian Mechanism and prove that it satisfies zCDP. 
We can then convert the privacy guarantees to approximate differential privacy.
We combine this with the \atd~ technique for a discrete variant of \Cref{alg:main}.
This approach does not give the tightest privacy parameters, but it is significantly simpler than a direct analysis of approximate differential privacy for multivariate discrete Gaussian noise~(see \cite[Theorem~2.14]{Canonne_Kamath_Steinke_2022}).
The primary contribution in this section is a simple change that makes the correlated Gaussian mechanism compatible with the discrete Gaussian.
We leave providing tighter analysis of the privacy guarantees open for future work.

\begin{algorithm}[H]
    \caption{The Discrete Correlated Gaussian Mechanism}
    \label{alg:discrete-correlated-gaussian}
    \begin{algorithmic}[1]
        \Require Parameters $k$ and $\sigma$, monotonic histogram $H(\X) \in \mathbb{N}^k$. 
        \State Let $\vec{\tilde{H}} = \{0\}^k$.
        \State Sample $Z_{\text{corr}} \sim \discGauss\left(0, 4\sigma^2/\sqrt{k}\right)$.
        \For{each $i \in [k]$}
        \State Sample $Z_i \sim \discGauss\left(0, 4\sigma^2\right)$.
        \State Set $\tilde{H}_i = H(\X)_i + (Z_i + Z_{\text{corr}})/2$.
        \EndFor
        \State \textbf{Release} $\tilde{H}$. 
    \end{algorithmic}
\end{algorithm}

\begin{lemma}
    \label{lem:discrete-cor-privacy}
    \Cref{alg:discrete-correlated-gaussian} satisfies $\rho$-zCDP where $\rho = (k + \sqrt{k})/(8\sigma^2)$.
\end{lemma}

\begin{proof}
    Fix any pair of neighboring histograms $H(\X)$ and $H(\X')$.
    We prove the lemma for the case where $H(\X) - H(\X') \in \{0,1\}^k$. 
    The proof is symmetric when $H(\X) - H(\X') \in \{0,-1\}^k$.

    Construct a new pair of histograms $\hat{H}(\X), \hat{H}(\X') \in \mathbb{N}^{k + 1}$ such that $\hat{H}(\X)_i = 2 \cdot H(\X)_i$ for all $i \in [k]$ and $\hat{H}(\X)_{k + 1} = 0$. 
    We set $\hat{H}(\X')_i = 2 \cdot \hat{H}(\X')_i - 1$ for all $i \in [k]$ and finally $\hat{H}(\X')_{k + 1} = 1$. 
    
    We clearly have that $\hat{H}(\X) \in \mathbb{Z}^{k + 1}$ for all possible $H(\X)$ as required by \Cref{lem:discrete-gaussian-mechanism}. 
    Now, we set $\sigma^2_i = 4\sigma^2$ for $i \in [k]$ and $\sigma^2_{k+1} = 4\sigma^2/\sqrt{k}$. 
    We constructed $\hat{H}(\X)$ and $\hat{H}(\X')$ such that they differ by $1$ in all entries for any pair of neighboring histograms. We therefore have that
    \[
        \sum_{i \in [k + 1]} (\hat{H}(\X)_i - \hat{H}(\X')_i)^2/\sigma_i^2 = \sum_{i \in [k + 1]} 1/\sigma_i^2 = k/(4\sigma^2) + \sqrt{k}/(4\sigma^2) = 2\rho \,.
    \]
    By \Cref{lem:discrete-gaussian-mechanism} we have that $D_\alpha(M(\X) || M(\X')) \leq \rho \alpha$ for all $\alpha > 1$, where $M(\X) = \hat{H}(\X) + Z$ and $Z_i \sim \Nd(0, \sigma_i^2)$.
    The output of $M(\X)$ can be post-processes such that $\tilde{H}_i = (M(\X)_i + M(\X)_{k+1})/2$.
    Notice that such a post-processing gives us the same output distribution as \Cref{alg:discrete-correlated-gaussian} for both $H(\X)$ and $H(\X')$.
    The algorithm therefore satisfies $\rho$-zCDP because the $\alpha$-Rényi divergence cannot increase from post-processing by \Cref{def:zero-concentrated-differential-privacy}.
\end{proof}

Note that the scaling step is crucial for the privacy proof.
It would not be sufficient to simply replace $Z_i \sim \Nd\left(0, \sigma^2\right)$ with $Z_i \sim \discGauss\left(0, \sigma^2\right)$.
We need to sample noise in discrete steps of length $1/2$ instead of length $1$.
Otherwise, the trick of centering the differences between the histograms using the $k+1$ entry as an offset does not work.
If we prefer a mechanism that always outputs integers, we can simply post-process the output.


\begin{lemma}
    \label{lem:discrete-upperbounddeltainf}
Let $Z_{\text{corr}} \sim \discGauss(0, 4\sigma^2/\sqrt{k})$ and $Z_1, \dots, Z_k \sim \discGauss(0, 4\sigma^2)$.
Then for any $\tau > 0$ we have that $\Pr[\exists i \in [k]: Z_{\text{corr}} + Z_i > 2\tau] \leq 1 - \Pr\left[\discGauss(0, 4\sigma^2/\sqrt{k}) \leq \dfrac{2\tau k^{-1/4}}{k^{-1/4}+1}\right] \cdot \Pr\left[\discGauss(0, 4\sigma^2) \leq \dfrac{2\tau}{k^{-1/4}+1}\right]^{k}$.
\end{lemma}

\begin{proof}
    The proof follows the same structure as for \Cref{claim:upperbounddeltainf}. 
    If $Z_{\text{corr}} \leq 2\tau k^{-1/4}/({k^{-1/4}+1})$ and $\max_{Z_1, \dots, Z_j} Z_i = \tilde{Z} \leq {2\tau}/({k^{-1/4}+1})$ then the sums cannot be above $2\tau$.
    The only difference is that we have to split up the probabilities. 
    In the discrete case the probabilities slightly change when we rescale.
\end{proof}

Finally, we can use replace the Gaussian noise in \Cref{alg:main} with discrete Gaussian noise.
Using the \atd~ approach and the lemmas above we get that.

\begin{theorem}
    \label{theorem:discrete-cor-GSHM}
    For parameters $k$, $\sigma$, and $\tau$, consider the mechanism $\mathcal{M} \colon \mathbb{N}^d \rightarrow \mathbb{Q}^d$ that given a histogram $H(\X)$: (1) runs \Cref{alg:discrete-correlated-gaussian} with parameters $k$ and $\sigma$ restricted to the support of $H(\X)$ (2) removes all noisy counts less than or equal to $1 + \tau$.
    Then $\mathcal{M}$ satisfies $(\varepsilon, \deltaone + \deltatwo)$-DP for $k$-sparse monotonic histograms, where 
    $\deltaone$ is such that $(k + \sqrt{k})/(8\sigma^2)$-zCDP implies $(\varepsilon, \deltaone)$-DP and
    \begin{align*}
    \deltatwo &= 1 - \Pr\left[\discGauss(0, 4\sigma^2/\sqrt{k}) \leq \dfrac{2 \tau k^{-1/4}}{k^{-1/4}+1}\right] \cdot \Pr\left[\discGauss(0, 4\sigma^2) \leq \dfrac{2 \tau}{k^{-1/4}+1}\right]^{k} \,.
    \end{align*}
\end{theorem}

\begin{proof}
    The proof relies on the construction of an intermediate histogram from the proof of \Cref{thm:algorithm-privacy-add-the-deltas}.
    We have that
    \begin{align*}
        \Pr[\mathcal{M}(H(\X)) \in Y] &\leq \Pr[\mathcal{M}(H(\hat{\X})) \in Y] + \deltatwo \\ 
        &\leq e^\varepsilon \Pr[\mathcal{M}(H(\X')) \in Y] + \deltaone  + \deltatwo \,,
    \end{align*}
    where the first inequality follows from \Cref{lem:discrete-upperbounddeltainf} and the second inequality follows from \Cref{lem:discrete-cor-privacy}.
    Both inequalities rely on the fact that the histograms are $k$-sparse.
\end{proof}

%% file: content/open.tex
\section{Conclusion and Open Problems}
\label{sec:conclusion}

We introduced the \emph{Correlated Stability Histogram} for the setting of $k$-sparse monotonic histograms and provided privacy guarantees using the \atd~approach and a more fine-grained case-by-case analysis.
We show that our mechanism outperforms the state-of-the-art -- the \emph{Gaussian Sparse Histogram Mechanism} -- and improves the utility by up to a factor of $2$.
In addition to various extensions, we enriched our theoretical contributions with a step towards practice by including a version that works with discrete Gaussian noise.

Unlike the previous work \cite{Wilkins24-GaussianSparseHistogramMechanism}, our bound in \cref{thm:privacy-tighter-analysis} is not tight.
It would be interesting to derive exact bounds for our mechanism as well.
Finally, we point out that the uncorrelated GSHM is still preferred in some settings.
The CSH requires an upper bound on $\|H(\X)\|_0$, whereas GSHM requires a bound on $\|X_i\|_0$.
If we have access to a bound $\|X_i\|_0 \leq k$ but no sparsity bound, we can apply our technique from \Cref{sec:top-k}.
This approach works well when the dataset is close to $k$-sparse, but if the dataset is large with many high counters the pre-processing step can introduce high error.
Furthermore, if histograms are $k$-sparse, but we additionally know that $\|X_i\|_0 \leq m$, for some $m < k$, our improvement factor changes.
If $m \leq k/4$, the GSHM has a lower error than the CSH.
However, in that setting, it might still be possible to reduce the error using the technique of \citet{JosephYu2024-constructions-k-norm-elliptic}.
We leave exploring this regime for future work.

%% file: content/ack.tex
\begin{acks}
Retschmeier carried out this work at Basic Algorithms Research Copenhagen (BARC), which was supported by the VILLUM Foundation grant 54451. 
\emph{Providentia}, a Data Science Distinguished Investigator grant from the Novo Nordisk Fonden, supported Retschmeier.
The work of Lebeda is supported by grant ANR-20-CE23-0015 (Project PRIDE).
\end{acks}

%% file: content/appendix.tex
\input{content/tables/table-of-symbols}

\section{Proof of \Cref{thm:privacy-tighter-analysis} (Tighter Analysis)}\label{app:proof-tighter-analysis}

We omitted the proof of \Cref{thm:privacy-tighter-analysis} from the main part of the paper.
Before providing the proof of theorem, we prove a useful lemma. 
We use this lemma in the proof of the theorem to bound the effect of the magnitude for the Gaussian noise on the value of $\delta$.

\begin{lemma}
    \label{lem:difference-for-j-sparse}
    Let $H(\X)$ and $H(\X')$ be a pair of $j$-sparse monotonic histograms with the same support.
    Then for \Cref{alg:main} with parameters $k \geq j$, $\sigma$ and $\tau$, we have
    \[
        \Pr[\algoname(H(\X)) \in Y] \leq e^\varepsilon \Pr[\algoname(H(\X')) \in Y] + \delta
    \].

    for any $\varepsilon \geq 0$ where $\gamma = \min(\dfrac{1}{2}\sqrt{j + \sqrt{k}}, \sqrt{j})$ and
    \[
    \delta \geq \Phi \left( \frac{\gamma}{2\sigma} - \frac{\varepsilon \sigma}{\gamma} \right) - e^\varepsilon \Phi \left( - \frac{\gamma}{2\sigma} - \frac{\varepsilon \sigma}{\gamma} \right) \,.
    \]
\end{lemma}

\begin{proof}
    We show that the statement holds for each case of the minimum for $\gamma$ separately.
    In each case, we show that the output distribution is equivalent to post-processing a Gaussian noise mechanism.
    We start with the case of $\gamma = \frac{1}{2}\sqrt{j + \sqrt{k}}$.
    Notice that by \Cref{lem:restated-Leb24-for-approx}, the inequality holds if we apply the correlated Gaussian mechanism to histograms with at most $j$ entries and set the magnitude of $Z_{cor}$ to $\sigma/\sqrt{k}$.
    Notice that the output distributions of $\algoname$ for both $H(\X)$ or $H(\X')$ as input is equivalent to running the Correlated Gaussian Mechanism restricted to their support followed by a post-processing step of removing values below $1 + \tau$. 

    Similarly, for the case of $\gamma = \sqrt{j}$ notice that $\|H(\X) - H(\X')\|_2 \leq \gamma$. 
    Therefore, we can output an $(\varepsilon, \delta)$-DP estimate of the histogram using the Standard Gaussian Mechanism with the parameters chosen according to the lemma.
    Now, notice that if we apply the Standard Gaussian Mechanism %
    to the histograms, we can recover the output distribution of $\algoname$ by post-processing. 
    We first sample noise $Z_{cor} \sim \Nd(0,\sigma/\sqrt{k})$ and add the value to all entries, and then removing all noise counts below $1 + \tau$ and any count not in the support of the histograms.

    The lemma thus holds since differential privacy is preserved under post-processing.
\end{proof}

\tighteranalysis*
\begin{proof}
    We prove the theorem by considering all values of the sensitivity space of $k$-sparse monotonic histograms.
    We start with the simplest cases. 
    If either $U = \emptyset$ or $U' = \emptyset$ we only have to consider the infinite privacy loss event. 
    This gives us the first term of the max by \Cref{claim:upperbounddeltainf}.
    Similarly, when $U = U'$, we get the second term of the max using the proof for $\delta_{\text{gauss}}$ in \Cref{thm:algorithm-privacy-add-the-deltas}.

    For the remaining cases, we have to consider both the impact of Gaussian noise and the potential for the infinite privacy loss event similar to \Cref{thm:algorithm-privacy-add-the-deltas}.
    We analyze the two cases of $U \subset U'$ and $U' \subset U$ separately.
    Without loss of generality, we assume that $\max(\vert U \vert, \vert U' \vert) = k$, since this is the worst-case for the privacy parameters.
    Furthermore, we have to analyze the mechanism when $\min(\vert U \vert, \vert U' \vert) = j$ for each possible $j \in [k - 1]$.

    First, we consider the case where $U' \subset U$.
    We construct an intermediate histogram $H(\hat{\X})$ as in the proof of \Cref{thm:algorithm-privacy-add-the-deltas}.
    The histogram $H(\hat{\X})$ has support $U'$, and $H(\hat{\X})_i = H(\X)_i$ for all $i \in U'$.
    Since $H(\hat{\X})$ and $H(\X')$ are a pair of $j$-sparse monotonic histograms we have by \Cref{lem:difference-for-j-sparse} that 
    \[
        \Pr[\algoname(H(\hat{\X})) \in Y] \leq e^\varepsilon \Pr[\algoname(H(\X')) \in Y] + \deltaone(j) \,,
    \]
    where 
    \[
    \deltaone(j) = \Phi \left( \frac{\gamma(j)}{2\sigma} - \frac{\varepsilon \sigma}{\gamma(j)} \right) - e^\varepsilon \Phi \left( - \frac{\gamma(j)}{2\sigma} - \frac{\varepsilon \sigma}{\gamma(j)} \right) \,.
    \]
    The histograms $H(\hat{\X})$ and $H(\X)$ only differs the entries $U \setminus U'$. Since $\vert U \setminus U' \vert = (k - j)$ we have that
    \[
        \Pr[\algoname(H(\X)) \in Y] \leq \Pr[\algoname(H(\hat{\X})) \in Y] + \deltatwo(j) \,,
    \]
    where $\deltatwo(j) = 1 - \psi({k-j})$ follows from \Cref{claim:upperbounddeltainf}.

    We recover the third term in the max using the values for $\deltatwo({j})$ and $\deltaone({j})$ and the fact that 
    \begin{align*}
        \Pr[\algoname(H(\X)) \in Y] &\leq \Pr[\algoname(H(\hat{\X})) \in Y] + \deltatwo(j) \\
        &\leq e^\varepsilon \Pr[\algoname(H(\X')) \in Y] + \deltaone(j) + \deltatwo(j) \,.
    \end{align*}

    Finally, we consider the case where $U \subset U'$. 
    This is the most technically involved case in the proof.
    Here, the intermediate histogram $H(\hat{\X})$ does not give us the bound we want because it is not $k$-sparse.
    Instead, we use an alternative representation of each histogram similar to the technique used in \cite{lebeda2024}. 
    We consider a mechanism for this representation and show that the bound holds when the alternative representations are used as input for this mechanism. 
    We then show that we can recover the distribution of \Cref{alg:main} using post-processing.
    We first give the proof when $\gamma = \frac{1}{2}\sqrt{j + \sqrt{k}}$ and then show that it also holds for $\gamma = \sqrt{j}$. %

    Assume without loss of generality that $U = [j]$ and $U' = [k]$ (we can always reorder the elements of the new representation, but this assumption makes the presentation simpler).
    We construct a vector $\vec{g} \in \mathbb{R}^{k+1}$ from $H(\X)$ such that $\vec{g}_i = H(\X)_i$ for each $i \in [j]$ and $\vec{g}_i = 0$ for any $i > j$. 
    For $\vec{g}' \in \mathbb{R}^{k+1}$ we set $\vec{g}'_i = H(\X')_i - 1/2$ for all $i \in [k]$ and we set $\vec{g}'_{k+1} = \frac{1}{2}k^{1/4}$.

    Now, we consider two mechanisms with input $\mathbb{R}^{k + 1}$.
    The first mechanism $\mathcal{G}$ ignore entries from index $j+1$ to $k$ and adds noise independently from $\mathcal{N}(0, \sigma^2)$ to all entries with index $i \in [j]$ as well as the $k + 1$'th entry.
    We want to show that $\Pr[\mathcal{G}(\vec{g}) \in Y] \leq e^{\hat{\varepsilon}(j)} \cdot \Pr[\mathcal{G}(\vec{g}') \in Y] + \hat{\delta}(j)$ for all $Y$ where 
    \[
        \hat{\delta}(j) = \Phi \left( \frac{\frac{1}{2}\sqrt{j + \sqrt{k}}}{2\sigma} - \frac{\hat{\varepsilon}(j) \sigma}{\frac{1}{2}\sqrt{j + \sqrt{k}}} \right) - e^{\hat{\varepsilon}(j)} \Phi \left( - \frac{\frac{1}{2}\sqrt{j + \sqrt{k}}}{2\sigma} - \frac{\hat{\varepsilon}(j) \sigma}{\frac{1}{2}\sqrt{j + \sqrt{k}}} \right) \,.
    \]
    Notice that $\mathcal{G}$ is equivalent to the standard Gaussian mechanism restricted to the domain $[j] \cup \{k + 1\}$. 
    At the same time, the distribution of $\mathcal{G}(\vec{g}')$ is equivalent to running $\mathcal{G}(\vec{\hat{g}})$ for a third vector $\vec{\hat{g}}$ where $\vec{\hat{g}}_i = \vec{\hat{g}}_i$ for $i \in ([j] \cup \{k + 1\})$ and $\vec{\hat{g}}_i = 0$ for $j < i \leq k$.
    The inequality above therefore follows from \Cref{lem:standard-gaussian-mechanism} if the $\ell_2$ distance between $\vec{g}$ and $\vec{\hat{g}}$ is at most $\frac{1}{2}\sqrt{j + \sqrt{k}}$ for any neighboring histograms.
    The absolute difference between $\vec{g}_i$ and $\vec{\hat{g}}_i$ is $1/2$ for each $i \in [j]$, while $\vec{g}_{k + 1}$ and $\vec{\hat{g}}_{k + 1}$ differ by $k^{1/4}/2$. 
    The $\ell_2$ distance between the vectors is therefore always
    $\|\vec{g} - \vec{\hat{g}}\|_2 = \left( \sum_{i \in [k + 1]} (\vec{g}_i - \vec{\hat{g}}_i)^2 \right)^{1/2} = \left( (k^{1/4}/2)^2 + \sum_{i \in [j]} (1/2)^2 \right)^{1/2} = \frac{1}{2}\sqrt{j + \sqrt{k}}$. 

    Next, consider a mechanism $\mathcal{F}$ that adds noise independently from $\mathcal{N}(0, \sigma^2)$ to all non-zero entries of a vector.
    Let $\vec{\tilde{g}}$ be the noisy vector after adding noise to a vector $\vec{g}$.
    If $\{j+1,\dots, k\}$ is in support of $\vec{g}$ we check if $\vec{\tilde{g}}_{k+1}/k^{1/4} + \vec{\tilde{g}}_i > \tau$ for any $i \in \{j+1,\dots, k\}$.
    In that case, we output $\bot$; otherwise, we set all such entries to $0$ and output the noisy vector $\vec{\tilde{g}}$.
    Here, we use the output $\bot$ to group all events corresponding to the infinite privacy loss in the original representation.

    We want to show that $\Pr[\mathcal{F}(\vec{g}) \in Y] \leq e^\varepsilon \Pr[\mathcal{F}(\vec{g}) \in Y] + \hat{\delta}(j)$ for all $Y$. 
    It is sufficient to show that the inequality holds for the set $Y_\varepsilon = \{y \mid e^\varepsilon \leq \mathcal{F}(\vec{g})(y)/\mathcal{F}(\vec{g'})(y)\}$, where $\mathcal{F}(\vec{g})(y)$ denotes the density at $y$ 
    (see \cite[Theorem~5]{Balle18-AnalyticalGaussian}).
    Now, if $\Pr[\mathcal{G}(\vec{g}') \in Y_\varepsilon] \leq \Pr[\mathcal{F}(\vec{g}') \in Y_\varepsilon]/\psi{(k-j)}$ the inequality holds because $\mathcal{G}(\vec{g})$ and $\mathcal{F}(\vec{g})$ has the same distribution and
    \begin{align*}
        \Pr[\mathcal{F}(\vec{g}) \in Y_\varepsilon] = \Pr[\mathcal{G}(\vec{g}) \in Y_\varepsilon] 
        & \leq e^{\hat{\varepsilon}} \Pr[\mathcal{G}(\vec{g}') \in Y_\varepsilon] + \hat{\delta}(j) \\
        & \leq e^{\varepsilon + \ln(\psi{(k-j)})} (\Pr[\mathcal{F}(\vec{g}') \in Y_\varepsilon]/\psi{(k-j)}) + \hat{\delta}(j) \\
        & = e^{\varepsilon} \Pr[\mathcal{F}(\vec{g}') \in Y_\varepsilon] + \hat{\delta}(j) \,, 
    \end{align*}
    where the last equality follows from $1/\psi{(k-j)} = e^{-\ln(\psi{(k-j)})}$.

    The missing piece above is to show that $\Pr[\mathcal{G}(\vec{g}') \in Y_\varepsilon] \leq \Pr[\mathcal{F}(\vec{g}') \in Y_\varepsilon]/\psi{(k-j)}$.
    The equivalent inequality is easy to see in the non-correlated variant of the Gaussian Sparse Histogram Mechanism because $Y_\varepsilon$ does not contain $\bot$ so $\Pr[\mathcal{F}'(\vec{g}') \in Y_\varepsilon] = \Pr[\mathcal{G}'(\vec{g}') \in Y_\varepsilon]\Pr[\top]$, where $\top$ is the complement of $\bot$, and $\mathcal{F}'$ and $\mathcal{G}'$ are the equivalents to $\mathcal{F}$ and $\mathcal{G}$ without the $(k+1)$'th entry. 
    The value of $\psi{(k-j)}$ is a lower bound on the probability of $\top$ occurring.
    Here we need to be a bit more careful here, but the inequality still holds because $\top$ occurring increases the probability that $\tilde{\vec{g}'}_{k+1}$ is small, which leads to a higher privacy loss event. 
    We sketch the argument below
    \begin{align*}
        \Pr[\mathcal{F}(\vec{g}') \in Y_\varepsilon] / \psi{(k-j)}
        & = \Pr[\mathcal{F}(\vec{g}') \in Y_\varepsilon \mid \top] \Pr[\top] / \psi{(k-j)} \\
        & \geq \Pr[\mathcal{F}(\vec{g}') \in Y_\varepsilon \mid \top] \geq \Pr[\mathcal{G}(\vec{g}') \in Y_\varepsilon] \,,
    \end{align*}
    where the last inequality relies on $\Pr[\mathcal{F}(\vec{g}') \in Y_\varepsilon \mid \tilde{\vec{g}}'_{k+1} \leq \beta \cap \top] = \Pr[\mathcal{G}(\vec{g}') \in Y_\varepsilon \mid \tilde{\vec{g}}'_{k+1} \leq \beta]$ for any $\beta \in \mathbb{R}$, 
    and $\Pr[\mathcal{G}(\vec{g}') \in Y_\varepsilon \mid \tilde{\vec{g}}'_{k+1} \leq \beta] \geq \Pr[\mathcal{G}(\vec{g}') \in Y_\varepsilon]$.
    For any $\beta$ we also have that $\Pr[\top \mid \mathcal{F}(\vec{g}') \leq \beta] \geq \Pr[\top]$ which allows us to apply Bayes' theorem to find that $\Pr[\mathcal{F}(\vec{g}')_{k+1} \leq \beta \mid \top] \geq \Pr[\mathcal{G}(\vec{g}')_{k+1} \leq \beta]$. 
    Integrating over all values of $\beta$ gives the inequality.
    
    Finally, we must show that \Cref{alg:main} is equivalent to post-processing $\mathcal{F}$. 
    Let $\vec{\tilde{g}}$ denote the output of $\mathcal{F}(\vec{g})$.
    Then we post-process the entries by first setting $\tilde{\vec{H}}_i = \vec{\tilde{g}}_i + \vec{\tilde{g}}_i/k^{1/4}$ and then removing all values of $\tilde{\vec{H}}_i$ below $1 + \tau$.
    If the output was $\bot$ we repeatable sample $\tilde{H} \sim \algoname(H(\X'))$ until $\tilde{H}_i \neq i$ for at least one $j < i \leq k$.
    We then output the histogram $\tilde{H}$.

    The proof is similar when $\gamma=\sqrt{j}$, but we use a different construction for $\vec{{g}}'$. 
    We set $\vec{{g}}'_i = H(\X')_i$ for all $i \in [k]$ and $\vec{{g}}'_{k + 1} = 0$.
    The inequality $\Pr[\mathcal{G}(\vec{g}) \in Y] \leq e^{\hat{\varepsilon}(j)} \cdot \Pr[\mathcal{G}(\vec{g}') \in Y] + \hat{\delta}(j)$ holds where $\frac{1}{2}\sqrt{j + \sqrt{k}}$ is replaced with $\sqrt{j}$ in $\hat{\delta}(j)$ %
    because $\|\vec{g} - \vec{\hat{g}}\|_2 = \left(\sum_{i \in [j]} (\vec{g}_i - \vec{\hat{g}}_i)^2\right)^{1/2} \leq \left(\sum_{i \in [j]} 1^2\right)^{1/2} = \sqrt{j}$.

    Since all possible elements in the sensitivity space falls in one of the categories above, we can compute the value of all inequalities above. 
    The maximum of all parameters gives us an upper bound for $\delta$.
\end{proof}

%% file: content/tables/table-of-symbols.tex
\section{Table of Symbols and Abbreviations}
The following \cref{tab:symbols} summarizes the symbols and abbreviations used throughout the paper.

\begin{table*}[h]
\centering
\resizebox{0.75\columnwidth}{!}{%
    \begin{tabular}{cl}
    {\bf Symbol} & {\bf Description} \\
    \hline
    GSHM & \textbf{G}aussian \textbf{S}parse \textbf{H}istogram \textbf{M}echanism\\
    \algoname & \textbf{C}orrelated \textbf{S}tability \textbf{H}istogram \\ %
    $\cA$ & Any aggregate query \\[0.65em]
    $\X, \tilde{\X}, \X' \in \mathcal{U}^{\N}$ & Datasets on universe $\mathcal{U}$\\
    $\X \sim \X'$ & Neighboring datasets\\
    $H(\X)$ & Histogram $H(\X) = \sum_i X_i$ \\
    $H(\X)^{(i)}$ & $i$'th largest item in $H(\X)$ \\
    $\|\vec{H}\|_0$ & $\ell_0$-norm of vector $\vec{H}$ (number of non-zeroes) \\
    $U, U'$ & Support of $H(\X),H(\X')$\\[0.65em]
    $\epsilon, \delta$ & Privacy Parameters \\
    $\tau > 0$ & Threshold is $1 + \tau$\\
    $\deltaone$ & $\delta$ from Gaussian noise\\
    $\deltatwo$; $\deltatwo^j$ & Probability of infinite privacy loss, for $j$ counts\\[0.75em]
    $\Nd(0, \sigma^2)$ & $0$-centered normal distribution\\
    $\discGauss(0, \sigma^2)$ & $0$-centered discrete normal distribution \cite{Canonne_Kamath_Steinke_2022} \\
    $\Phi(x); \Phi^{-1}(x)$ & CDF of the normal distribution; inverse CDF\\
    \hline
    \end{tabular}}\\
\caption{Symbols used throughout the paper}
\label{tab:symbols}
\end{table*}

%% file: main.bbl

\begin{thebibliography}{25}


\ifx \showCODEN    \undefined \def \showCODEN     #1{\unskip}     \fi
\ifx \showDOI      \undefined \def \showDOI       #1{#1}\fi
\ifx \showISBNx    \undefined \def \showISBNx     #1{\unskip}     \fi
\ifx \showISBNxiii \undefined \def \showISBNxiii  #1{\unskip}     \fi
\ifx \showISSN     \undefined \def \showISSN      #1{\unskip}     \fi
\ifx \showLCCN     \undefined \def \showLCCN      #1{\unskip}     \fi
\ifx \shownote     \undefined \def \shownote      #1{#1}          \fi
\ifx \showarticletitle \undefined \def \showarticletitle #1{#1}   \fi
\ifx \showURL      \undefined \def \showURL       {\relax}        \fi
\providecommand\bibfield[2]{#2}
\providecommand\bibinfo[2]{#2}
\providecommand\natexlab[1]{#1}
\providecommand\showeprint[2][]{arXiv:#2}

\bibitem[Aum{\"{u}}ller et~al\mbox{.}(2022)]%
        {ALP}
\bibfield{author}{\bibinfo{person}{Martin Aum{\"{u}}ller}, \bibinfo{person}{Christian~Janos Lebeda}, {and} \bibinfo{person}{Rasmus Pagh}.} \bibinfo{year}{2022}\natexlab{}.
\newblock \showarticletitle{Representing Sparse Vectors with Differential Privacy, Low Error, Optimal Space, and Fast Access}.
\newblock \bibinfo{journal}{\emph{Journal of Privacy and Confidentiality}} \bibinfo{volume}{12}, \bibinfo{number}{2} (\bibinfo{date}{Nov.} \bibinfo{year}{2022}).
\newblock
\urldef\tempurl%
\url{https://doi.org/10.29012/jpc.809}
\showDOI{\tempurl}


\bibitem[Bafna and Ullman(2017)]%
        {bafna2017price}
\bibfield{author}{\bibinfo{person}{Mitali Bafna} {and} \bibinfo{person}{Jonathan Ullman}.} \bibinfo{year}{2017}\natexlab{}.
\newblock \showarticletitle{The price of selection in differential privacy}. In \bibinfo{booktitle}{\emph{Conference on Learning Theory}}. PMLR, \bibinfo{publisher}{PMLR}, \bibinfo{pages}{151--168}.
\newblock


\bibitem[Balcer and Vadhan(2019)]%
        {BalcerVadhan-DP-finite-computers}
\bibfield{author}{\bibinfo{person}{Victor Balcer} {and} \bibinfo{person}{Salil Vadhan}.} \bibinfo{year}{2019}\natexlab{}.
\newblock \showarticletitle{Differential Privacy on Finite Computers}.
\newblock \bibinfo{journal}{\emph{Journal of Privacy and Confidentiality}} \bibinfo{volume}{9}, \bibinfo{number}{2} (\bibinfo{date}{Sep.} \bibinfo{year}{2019}).
\newblock
\urldef\tempurl%
\url{https://doi.org/10.29012/jpc.679}
\showDOI{\tempurl}


\bibitem[Balle and Wang(2018)]%
        {Balle18-AnalyticalGaussian}
\bibfield{author}{\bibinfo{person}{Borja Balle} {and} \bibinfo{person}{Yu{-}Xiang Wang}.} \bibinfo{year}{2018}\natexlab{}.
\newblock \showarticletitle{Improving the Gaussian Mechanism for Differential Privacy: Analytical Calibration and Optimal Denoising}. In \bibinfo{booktitle}{\emph{{ICML}}} \emph{(\bibinfo{series}{Proceedings of Machine Learning Research}, Vol.~\bibinfo{volume}{80})}. \bibinfo{publisher}{{PMLR}}, \bibinfo{pages}{403--412}.
\newblock


\bibitem[Bun et~al\mbox{.}(2016)]%
        {BunNS16-simultaneous-learning-multiple-concepts}
\bibfield{author}{\bibinfo{person}{Mark Bun}, \bibinfo{person}{Kobbi Nissim}, {and} \bibinfo{person}{Uri Stemmer}.} \bibinfo{year}{2016}\natexlab{}.
\newblock \showarticletitle{Simultaneous Private Learning of Multiple Concepts}. In \bibinfo{booktitle}{\emph{Proceedings of the 2016 {ACM} Conference on Innovations in Theoretical Computer Science, Cambridge, MA, USA, January 14-16, 2016}}, \bibfield{editor}{\bibinfo{person}{Madhu Sudan}} (Ed.). \bibinfo{publisher}{{ACM}}, \bibinfo{pages}{369--380}.
\newblock
\urldef\tempurl%
\url{https://doi.org/10.1145/2840728.2840747}
\showDOI{\tempurl}


\bibitem[Bun and Steinke(2016)]%
        {zeroConcentrated}
\bibfield{author}{\bibinfo{person}{Mark Bun} {and} \bibinfo{person}{Thomas Steinke}.} \bibinfo{year}{2016}\natexlab{}.
\newblock \showarticletitle{Concentrated differential privacy: Simplifications, extensions, and lower bounds}. In \bibinfo{booktitle}{\emph{Theory of Cryptography Conference}}. Springer, \bibinfo{pages}{635--658}.
\newblock


\bibitem[Canonne et~al\mbox{.}(2022)]%
        {Canonne_Kamath_Steinke_2022}
\bibfield{author}{\bibinfo{person}{Clement Canonne}, \bibinfo{person}{Gautam Kamath}, {and} \bibinfo{person}{Thomas Steinke}.} \bibinfo{year}{2022}\natexlab{}.
\newblock \showarticletitle{The Discrete Gaussian for Differential Privacy}.
\newblock \bibinfo{journal}{\emph{Journal of Privacy and Confidentiality}} \bibinfo{volume}{12}, \bibinfo{number}{1} (\bibinfo{date}{Jul.} \bibinfo{year}{2022}).
\newblock
\urldef\tempurl%
\url{https://doi.org/10.29012/jpc.784}
\showDOI{\tempurl}


\bibitem[Durfee and Rogers(2019)]%
        {durfee2019}
\bibfield{author}{\bibinfo{person}{David Durfee} {and} \bibinfo{person}{Ryan~M Rogers}.} \bibinfo{year}{2019}\natexlab{}.
\newblock \bibinfo{booktitle}{\emph{Practical differentially private top-k selection with pay-what-you-get composition}}. Vol.~\bibinfo{volume}{32}.
\newblock \bibinfo{publisher}{Curran Associates Inc.}, \bibinfo{address}{Red Hook, NY, USA}.
\newblock


\bibitem[Dwork et~al\mbox{.}(2006a)]%
        {DworkKMMN06OurDataOurselves}
\bibfield{author}{\bibinfo{person}{Cynthia Dwork}, \bibinfo{person}{Krishnaram Kenthapadi}, \bibinfo{person}{Frank McSherry}, \bibinfo{person}{Ilya Mironov}, {and} \bibinfo{person}{Moni Naor}.} \bibinfo{year}{2006}\natexlab{a}.
\newblock \showarticletitle{Our Data, Ourselves: Privacy Via Distributed Noise Generation}. In \bibinfo{booktitle}{\emph{Advances in Cryptology - {EUROCRYPT} 2006, 25th Annual International Conference on the Theory and Applications of Cryptographic Techniques, St. Petersburg, Russia, May 28 - June 1, 2006, Proceedings}} \emph{(\bibinfo{series}{Lecture Notes in Computer Science}, Vol.~\bibinfo{volume}{4004})}. \bibinfo{publisher}{Springer}, \bibinfo{pages}{486--503}.
\newblock
\urldef\tempurl%
\url{https://doi.org/10.1007/11761679\_29}
\showDOI{\tempurl}


\bibitem[Dwork et~al\mbox{.}(2006b)]%
        {dwork06calibrating}
\bibfield{author}{\bibinfo{person}{Cynthia Dwork}, \bibinfo{person}{Frank McSherry}, \bibinfo{person}{Kobbi Nissim}, {and} \bibinfo{person}{Adam Smith}.} \bibinfo{year}{2006}\natexlab{b}.
\newblock \showarticletitle{Calibrating noise to sensitivity in private data analysis}. In \bibinfo{booktitle}{\emph{Theory of Cryptography: Third Theory of Cryptography Conference, TCC 2006, New York, NY, USA, March 4-7, 2006. Proceedings 3}}. Springer, \bibinfo{pages}{265--284}.
\newblock


\bibitem[Dwork and Roth(2014)]%
        {dworkRothBook}
\bibfield{author}{\bibinfo{person}{Cynthia Dwork} {and} \bibinfo{person}{Aaron Roth}.} \bibinfo{year}{2014}\natexlab{}.
\newblock \showarticletitle{The Algorithmic Foundations of Differential Privacy}.
\newblock \bibinfo{journal}{\emph{Foundations and Trends in Theoretical Computer Science}} \bibinfo{volume}{9}, \bibinfo{number}{3-4} (\bibinfo{year}{2014}), \bibinfo{pages}{211--407}.
\newblock
\urldef\tempurl%
\url{https://doi.org/10.1561/0400000042}
\showDOI{\tempurl}


\bibitem[{\relax Google Anonymization Team}(2020)]%
        {googlelibthreshold}
\bibfield{author}{\bibinfo{person}{{\relax Google Anonymization Team}}.} \bibinfo{year}{2020}\natexlab{}.
\newblock \bibinfo{title}{Delta for thresholding}.
\newblock \bibinfo{howpublished}{\url{https://github.com/google/differential-privacy/blob/main/common_docs/Delta_For_Thresholding.pdf}}.
\newblock
\newblock
\shownote{[Online; accessed 8-December-2024]}.


\bibitem[Gotz et~al\mbox{.}(2012)]%
        {Gotz2012}
\bibfield{author}{\bibinfo{person}{Michaela Gotz}, \bibinfo{person}{Ashwin Machanavajjhala}, \bibinfo{person}{Guozhang Wang}, \bibinfo{person}{Xiaokui Xiao}, {and} \bibinfo{person}{Johannes Gehrke}.} \bibinfo{year}{2012}\natexlab{}.
\newblock \showarticletitle{Publishing Search Logs—A Comparative Study of Privacy Guarantees}.
\newblock \bibinfo{journal}{\emph{IEEE Transactions on Knowledge and Data Engineering}} \bibinfo{volume}{24}, \bibinfo{number}{3} (\bibinfo{year}{2012}), \bibinfo{pages}{520--532}.
\newblock
\urldef\tempurl%
\url{https://doi.org/10.1109/TKDE.2011.26}
\showDOI{\tempurl}


\bibitem[Joseph and Yu(2024)]%
        {JosephYu2024-constructions-k-norm-elliptic}
\bibfield{author}{\bibinfo{person}{Matthew Joseph} {and} \bibinfo{person}{Alexander Yu}.} \bibinfo{year}{2024}\natexlab{}.
\newblock \showarticletitle{Some Constructions of Private, Efficient, and Optimal K-Norm and Elliptic Gaussian Noise}. In \bibinfo{booktitle}{\emph{The Thirty Seventh Annual Conference on Learning Theory, June 30 - July 3, 2023, Edmonton, Canada}} \emph{(\bibinfo{series}{Proceedings of Machine Learning Research}, Vol.~\bibinfo{volume}{247})}. \bibinfo{publisher}{{PMLR}}, \bibinfo{pages}{2723--2766}.
\newblock
\urldef\tempurl%
\url{https://proceedings.mlr.press/v247/joseph24a.html}
\showURL{%
\tempurl}


\bibitem[Korolova et~al\mbox{.}(2009)]%
        {Korolova09-DP-approx-sparse-hist}
\bibfield{author}{\bibinfo{person}{Aleksandra Korolova}, \bibinfo{person}{Krishnaram Kenthapadi}, \bibinfo{person}{Nina Mishra}, {and} \bibinfo{person}{Alexandros Ntoulas}.} \bibinfo{year}{2009}\natexlab{}.
\newblock \showarticletitle{Releasing search queries and clicks privately}. In \bibinfo{booktitle}{\emph{{WWW}}}. \bibinfo{publisher}{{ACM}}, \bibinfo{pages}{171--180}.
\newblock
\urldef\tempurl%
\url{https://doi.org/10.1145/1526709.1526733}
\showDOI{\tempurl}


\bibitem[Lebeda(2024)]%
        {lebeda2024}
\bibfield{author}{\bibinfo{person}{Christian~Janos Lebeda}.} \bibinfo{year}{2024}\natexlab{}.
\newblock \bibinfo{title}{Better Gaussian Mechanism using Correlated Noise}.
\newblock
\newblock
\showeprint[arxiv]{2408.06853}~[cs.CR]
\urldef\tempurl%
\url{https://arxiv.org/abs/2408.06853}
\showURL{%
\tempurl}


\bibitem[Lebeda and Tetek(2023)]%
        {LebedaTetek23-DPMG}
\bibfield{author}{\bibinfo{person}{Christian~Janos Lebeda} {and} \bibinfo{person}{Jakub Tetek}.} \bibinfo{year}{2023}\natexlab{}.
\newblock \showarticletitle{Better Differentially Private Approximate Histograms and Heavy Hitters using the Misra-Gries Sketch}. In \bibinfo{booktitle}{\emph{{PODS}}}. \bibinfo{publisher}{{ACM}}, \bibinfo{pages}{79--88}.
\newblock


\bibitem[McSherry and Talwar(2007)]%
        {mcsherry2017}
\bibfield{author}{\bibinfo{person}{Frank McSherry} {and} \bibinfo{person}{Kunal Talwar}.} \bibinfo{year}{2007}\natexlab{}.
\newblock \showarticletitle{Mechanism design via differential privacy}. In \bibinfo{booktitle}{\emph{48th Annual IEEE Symposium on Foundations of Computer Science (FOCS'07)}}. IEEE, \bibinfo{publisher}{IEEE}, \bibinfo{pages}{94--103}.
\newblock


\bibitem[Messing et~al\mbox{.}(2020)]%
        {DVN/TDOAPG_2020}
\bibfield{author}{\bibinfo{person}{Solomon Messing}, \bibinfo{person}{Christina DeGregorio}, \bibinfo{person}{Bennett Hillenbrand}, \bibinfo{person}{Gary King}, \bibinfo{person}{Saurav Mahanti}, \bibinfo{person}{Zagreb Mukerjee}, \bibinfo{person}{Chaya Nayak}, \bibinfo{person}{Nate Persily}, \bibinfo{person}{Bogdan State}, {and} \bibinfo{person}{Arjun Wilkins}.} \bibinfo{year}{2020}\natexlab{}.
\newblock \bibinfo{title}{{Facebook Privacy-Protected Full URLs Data Set}}.
\newblock
\newblock
\urldef\tempurl%
\url{https://doi.org/10.7910/DVN/TDOAPG}
\showDOI{\tempurl}


\bibitem[Misra and Gries(1982)]%
        {MisraGries82}
\bibfield{author}{\bibinfo{person}{J. Misra} {and} \bibinfo{person}{David Gries}.} \bibinfo{year}{1982}\natexlab{}.
\newblock \showarticletitle{Finding repeated elements}.
\newblock \bibinfo{journal}{\emph{Science of Computer Programming}} \bibinfo{volume}{2}, \bibinfo{number}{2} (\bibinfo{year}{1982}), \bibinfo{pages}{143--152}.
\newblock
\showISSN{0167-6423}
\urldef\tempurl%
\url{https://doi.org/10.1016/0167-6423(82)90012-0}
\showDOI{\tempurl}


\bibitem[Qiao et~al\mbox{.}(2021)]%
        {QiaoSZ21}
\bibfield{author}{\bibinfo{person}{Gang Qiao}, \bibinfo{person}{Weijie~J. Su}, {and} \bibinfo{person}{Li Zhang}.} \bibinfo{year}{2021}\natexlab{}.
\newblock \showarticletitle{Oneshot Differentially Private Top-k Selection}. In \bibinfo{booktitle}{\emph{Proceedings of the 38th International Conference on Machine Learning, {ICML} 2021, 18-24 July 2021, Virtual Event}} \emph{(\bibinfo{series}{Proceedings of Machine Learning Research}, Vol.~\bibinfo{volume}{139})}, \bibfield{editor}{\bibinfo{person}{Marina Meila} {and} \bibinfo{person}{Tong Zhang}} (Eds.). \bibinfo{publisher}{{PMLR}}, \bibinfo{pages}{8672--8681}.
\newblock


\bibitem[Rosén(1997)]%
        {ROSEN1997135}
\bibfield{author}{\bibinfo{person}{Bengt Rosén}.} \bibinfo{year}{1997}\natexlab{}.
\newblock \showarticletitle{Asymptotic theory for order sampling}.
\newblock \bibinfo{journal}{\emph{Journal of Statistical Planning and Inference}} \bibinfo{volume}{62}, \bibinfo{number}{2} (\bibinfo{year}{1997}), \bibinfo{pages}{135--158}.
\newblock
\showISSN{0378-3758}
\urldef\tempurl%
\url{https://doi.org/10.1016/S0378-3758(96)00185-1}
\showDOI{\tempurl}


\bibitem[Wilkins et~al\mbox{.}(2024)]%
        {Wilkins24-GaussianSparseHistogramMechanism}
\bibfield{author}{\bibinfo{person}{Arjun Wilkins}, \bibinfo{person}{Daniel Kifer}, \bibinfo{person}{Danfeng Zhang}, {and} \bibinfo{person}{Brian Karrer}.} \bibinfo{year}{2024}\natexlab{}.
\newblock \showarticletitle{Exact Privacy Analysis of the Gaussian Sparse Histogram Mechanism}.
\newblock \bibinfo{journal}{\emph{Journal of Privacy and Confidentiality}} \bibinfo{volume}{14}, \bibinfo{number}{1} (\bibinfo{date}{Feb.} \bibinfo{year}{2024}).
\newblock
\urldef\tempurl%
\url{https://doi.org/10.29012/jpc.823}
\showDOI{\tempurl}


\bibitem[Wilson et~al\mbox{.}(2020)]%
        {WilsonZLDSG20-SQL-bounded-contribution}
\bibfield{author}{\bibinfo{person}{Royce~J. Wilson}, \bibinfo{person}{Celia~Yuxin Zhang}, \bibinfo{person}{William Lam}, \bibinfo{person}{Damien Desfontaines}, \bibinfo{person}{Daniel Simmons{-}Marengo}, {and} \bibinfo{person}{Bryant Gipson}.} \bibinfo{year}{2020}\natexlab{}.
\newblock \showarticletitle{Differentially Private {SQL} with Bounded User Contribution}.
\newblock \bibinfo{journal}{\emph{Proc. Priv. Enhancing Technol.}} \bibinfo{volume}{2020}, \bibinfo{number}{2} (\bibinfo{year}{2020}), \bibinfo{pages}{230--250}.
\newblock
\urldef\tempurl%
\url{https://doi.org/10.2478/POPETS-2020-0025}
\showDOI{\tempurl}


\bibitem[Wu and Zhang(2024)]%
        {hao2024}
\bibfield{author}{\bibinfo{person}{Hao Wu} {and} \bibinfo{person}{Hanwen Zhang}.} \bibinfo{year}{2024}\natexlab{}.
\newblock \showarticletitle{Faster Differentially Private Top-$k$ Selection: A Joint Exponential Mechanism with Pruning}. In \bibinfo{booktitle}{\emph{Proceedings of the 37th International Conference on Neural Information Processing Systems}}. NeurIPS, \bibinfo{publisher}{NeurIPS}.
\newblock


\end{thebibliography}
